\documentstyle[12pt]{article}

\newcommand{\eqn}[1]{(\ref{#1})}
\newcommand{\complex}{{\bb C}} 
\newcommand{\zed}{{\bb Z}} 
\newcommand{\IR}{{\bb R}} 
\newcommand{\id}{{1\!\!1}} 
\newcommand{\orbits}{{\bbs M}} 
\newcommand{\orbit}{{\bb M}} 
\newcommand{\MDD}{{\cal M}_{d+1}}  
\newcommand{\MD}{{\cal M}_d}  

\font\mybb=msbm10 at 12pt
\def\bb#1{\hbox{\mybb#1}}
\font\mybbs=msbm10 at 9pt
\def\bbs#1{\hbox{\mybbs#1}}

\def\e{{\rm e}}
\def\beq{\begin{equation}}
\def\eeq{\end{equation}}
\def\bea{\begin{eqnarray}}
\def\eea{\end{eqnarray}}
\newcommand{\nn}{\nonumber}
\def\bd{\begin{displaymath}}
\def\ed{\end{displaymath}}

\makeatletter
\newdimen\normalarrayskip              
\newdimen\minarrayskip                 
\normalarrayskip\baselineskip
\minarrayskip\jot
\newif\ifold             \oldtrue            \def\new{\oldfalse}
\def\arraymode{\ifold\relax\else\displaystyle\fi} 
\def\@arrayskip{\ifold\baselineskip\z@\lineskip\z@
     \else
     \baselineskip\minarrayskip\lineskip2\minarrayskip\fi}
\def\@arrayclassz{\ifcase \@lastchclass \@acolampacol \or
\@ampacol \or \or \or \@addamp \or
   \@acolampacol \or \@firstampfalse \@acol \fi
\edef\@preamble{\@preamble
  \ifcase \@chnum
     \hfil$\relax\arraymode\@sharp$\hfil
     \or $\relax\arraymode\@sharp$\hfil
     \or \hfil$\relax\arraymode\@sharp$\fi}}
\def\@array[#1]#2{\setbox\@arstrutbox=\hbox{\vrule
     height\arraystretch \ht\strutbox
     depth\arraystretch \dp\strutbox
     width\z@}\@mkpream{#2}\edef\@preamble{\halign \noexpand\@halignto
\bgroup \tabskip\z@ \@arstrut \@preamble \tabskip\z@ \cr}%
\let\@startpbox\@@startpbox \let\@endpbox\@@endpbox
  \if #1t\vtop \else \if#1b\vbox \else \vcenter \fi\fi
  \bgroup \let\par\relax
  \let\@sharp##\let\protect\relax
  \@arrayskip\@preamble}
\makeatother

\newcommand{\newsection}[1]
{\vspace{5mm}
\pagebreak[3]
\addtocounter{section}{1}
\setcounter{equation}{0}
\setcounter{subsection}{0}
\setcounter{footnote}{0}
\begin{flushleft}
{\large\bf \thesection. #1}
\end{flushleft}
\nopagebreak
\medskip
\nopagebreak}

\newcommand{\newsubsection}[1]{
 \vspace{5mm}
\pagebreak[3]
\addtocounter{subsection}{1}
\noindent{ \bf \thesubsection. #1}
\nopagebreak
\vspace{2mm}
\nopagebreak}

\newlength{\extraspace}
\newlength{\extraspaces}
\begin{document}

\renewcommand{\footnotesize}{\small}

\addtolength{\baselineskip}{.8mm}

\thispagestyle{empty}

\begin{flushright}
\baselineskip=12pt
NBI-HE-99-21\\
hep-th/9908051\\
\hfill{  }\\ August 1999
\end{flushright}
\vspace{.2cm}

\begin{center}

\baselineskip=24pt

{\Large\bf Topological Field Theory and
Quantum Holonomy Representations of Motion Groups}\\[15mm]

\baselineskip=12pt

{\bf Richard J. Szabo}\\[3mm]
{\it Department of Physics and Astronomy, University of British Columbia\\
6224 Agricultural Road, Vancouver, B.C. V6T 1Z1, Canada}\\[2mm]
and\\[2mm]
{\it The Niels Bohr Institute\\ Blegdamsvej 17, DK-2100 Copenhagen \O,
Denmark}\\[2mm] {\tt szabo@nbi.dk}\\[30mm]

{\sc Abstract}

\begin{center}
\begin{minipage}{14cm}

\baselineskip=12pt

Canonical quantization of abelian $BF$-type topological field theory 
coupled to extended sources on generic
$d$-dimensional manifolds and with curved line bundles is
studied. Sheaf cohomology is used to construct the appropriate topological
extension of the action and the topological flux quantization conditions,
in terms of the \v{C}ech cohomology of the underlying spatial manifold, as required for
topological invariance of the quantum field theory. The wavefunctions
are found in the Hamiltonian formalism and are shown to carry multi-dimensional
projective representations of various topological groups of the space.
Expressions for generalized linking numbers in any dimension are thereby
derived.
In particular, new global aspects of motion group presentations are obtained
in any dimension. Applications to quantum exchange statistics of objects in
various dimensionalities are also discussed.

\end{minipage}
\end{center}

\end{center}

\vfill
\newpage
\pagestyle{plain}
\setcounter{page}{1}
\stepcounter{subsection}

\newsection{Introduction}

Topological gauge theories involving higher-rank antisymmetric tensor fields
have been of much interest over the years. The simplest example of such a model
is $BF$ theory \cite{schwarz}--\cite{bt} which
provides a quantum field theoretical
framework for understanding various important topological
invariants. In the abelian case,
the partition function computes
the Ray-Singer analytic torsion of the underlying spacetime 
manifold \cite{schwarz,bt,gk} while
the correlation functions of higher-rank holonomy operators (the appropriate 
generalizations of Wilson loop operators) compute linking and intersection
numbers of manifolds of various dimensionalities
\cite{bt,hs}. These models have also been of
interest in a wide variety of physical
applications in which non-local holonomy effects arising
from processes involving adiabatic
transports of extended
objects, such as strings, play a significant role \cite{hol}.   
Recent interest in non-abelian $BF$ theory has been sparked by the realization
that it serves as a dual model for Yang-Mills theory and quantum
chromodynamics in four dimensions \cite{YM} and thereby provides a computational
tool in which the non-perturbative sector of the relevant quantum field
theory can be analysed quantitatively. $BF$ theory has also played a 
significant role in various models of low dimensional quantum gravity 
\cite{qg}. An introduction to the applications of $BF$ theory in four dimensional
quantum gravity may be found in \cite{baez}, while a general concise introduction
to topological quantum field theory is given in \cite{tqft}.

In this paper we will study a general class of abelian topological $BF$ theories and
use it to describe some topological invariants of manifolds of generic
dimension. The basic mathematical motivation comes from a well-known 
classification problem in geometric topology. In the classification theory of
three-dimensional manifolds ${\cal M}_3$, 
an important invariant is the topological class
of a mapping of the circle ${\bf S}^1$ into ${\cal M}_3$ such that no two
points of the loop intersect on ${\cal M}_3$. The study of all
such embeddings is known as knot theory. Knots in dimensions other than three
are always trivial, but in three dimensions there is a vast collection of
such topological classes which live in the fundamental homotopy group
$\pi_1({\cal M}_3)$ of the manifold. An algebraic entity related to knot theory
is the braid group of a two-dimensional space \cite{braids,ladeg}.
A braid of $N$ strands can
be viewed as a collection of overlapping lines on the plane. It is a 
fundamental theorem of geometric topology that any collection of knots
or links can be constructed by joining the ends of a certain braid. The
braid group is therefore a useful tool for the classification of three-manifolds.
These structures can all be generalized to higher dimensional embeddings into
higher dimensional manifolds to form objects known as motion groups
\cite{dahm,goldsmith} which are the appropriate generalizations of braid
groups. The motion group is a useful tool in the topological classification of manifolds
whose properties have only been touched upon briefly in the mathematics
literature.

Polynomial invariants of knots in three dimensions and of two-knots in four
dimensions are derived in \cite{crm,ccr} from observables in various versions
of abelian and non-abelian $BF$ field theories.
In the following we will obtain a special class of 
multi-dimensional abelian unitary representations
of generic motion groups in manifolds of generic dimension. This
analysis sheds some light on the global structures involved in the
algebraic and geometric definitions of motion groups in general. Just as the braid
group and its representations via Chern-Simons gauge theory \cite{djt}--\cite{bs}
arise from the statistical
exchange holonomies between quantum mechanical point particles in two spatial
dimensions, the quantum holonomies induced in the wavefunctions for $BF$
gauge fields coupled minimally to higher dimensional worldvolumes lead to the
appropriate motion group representations and are relevant to the description 
of exchange statistics between extended objects in higher spatial dimensions.
The possibility of exotic statistics between strings in three dimensions
was considered in several different contexts in \cite{hol} and analysed in
some more generality in \cite{bmosss}. Field theoretical models using
four dimensional $BF$ theory were described in \cite{abks}--\cite{me} and
related to features of the motion group. The class of quantum field theories 
analysed in this paper is relevant to more general models of D-branes
and M-branes which have played a fundamental role in the present understanding
of the dynamics of superstring theory and M-theory. The results obtained
in the following describe various geometrical aspects of antisymmetric tensor fields
with non-trivial topological charges and present an essential description
of the quantum field theory for these extended objects.

We will be particularly interested in global aspects of these topological
field theories and their associated motion group representations. The basic
modifications which arise when the line bundle of the
gauge theory is topologically non-trivial can be understood easiest in
terms of the global gauge group of the $BF$ theory. For a $BF$ field theory
constructed from a $p$-form field and a corresponding dual $d-p-1$-form field
which are sections of a trivial vector bundle over a $d$-dimensional
manifold $\MD$, the harmonic zero mode contributions of the fields to the
corresponding path integral produce a Grassmann parity graded direct sum of deRham
cohomology groups \cite{bt},
\beq
{\cal Z}(\MD)=\bigoplus_{n=0}^{2p-d}(-1)^n\,H_{\rm D}^{p-n}(\MD)\oplus
\left(1+(-1)^{d+1}\right)\bigoplus_{m=2p-d+1}^p(-1)^m\,H_{\rm D}^{p-m}
(\MD)
\label{zerospace}\eeq
The appropriate modification for curved vector bundles
of the treatment of the harmonic zero modes
is described in general in \cite{cmr} and applied to abelian $BF$ theory
in \cite{newbf}. It is found there that 
the zero mode space \eqn{zerospace} is modified 
by the \v{C}ech cohomology of the manifold $\MD$. In the following we will
describe this modification within the framework of canonical quantization.
We shall use sheaf cohomology in a rather straightforward way \cite{bs,cechcs}
that is much simpler than the covariant approach of \cite{cmr} which
requires the use of \v{C}ech hyper-cohomology complexes and higher rank bundles
whose properties are not very well-understood except in some lower dimensional 
cases \cite{fnn,brylinski}. As we will show, the present sheaf cohomological
approach is particularly well-suited for the canonical quantization of the
source-coupled $BF$ theory, but is extremely cumbersome for a path integral
treatment along the lines of \cite{cmr,newbf}. Conversely, the former approach
is not very useful in a canonical framework because the various
degrees of freedom in the decompositions of the fields are not very well
understood. Having obtained the appropriate modification of the $BF$ action
which incorporates global effects, we will then be able to deduce the 
corresponding constraints that the generators of the motion group on a
homologically non-trivial manifold must satisfy.

The outline of this paper is as follows. In section 2 we describe,
using sheaf cohomology, the 
modifications of the $BF$ field theory action that must be made when the
relevant gauge fields are sections of a non-trivial line bundle. 
We show how this modification simplifies within a canonical framework
and also derive a topological quantization constraint on the fluxes of the 
gauge fields in terms of the \v{C}ech cohomology of the manifold which ensures
that the quantum field theory is indeed topologically invariant. 
In section 3 we introduce the parametrizations of the fields that will
be used and study the canonical structure of the field theory. We illustrate
here how the appropriate modification to \eqn{zerospace} arises in
the canonical formalism through strong gauge constraints which relate
the global fluxes of the gauge fields to the external charges of the
theory. In section 4 we solve explicitly for the wavefunctions of the
quantum field theory in the functional Schr\"odinger representation
and show how the sheaf cohomological constraints ensure
that they are indeed topological invariants. One important ingredient in
this construction is the derivation of a generalized solid angle function
which computes the adiabatic linking numbers of embedded surfaces of
generic dimension. In section 5 we introduce the motion group and describe
some of its formal mathematical properties, and then proceed to describe
the holonomy representations of it carried by the wavefunctions of the 
canonical quantum field theory. Section 6 then concludes with some
examples, applications and directions for further generalizations. We
show how the standard braid group representations induced from
Chern-Simons gauge theory on a Riemann surface \cite{bosnair,bs} are recovered
within the present formalism but with some global modifications. These
global modifications are also shown to occur for the quantum exchange holonomies
of equal rank extended objects in higher dimensions. The main feature is that 
the $BF$ formalism avoids the cumbersome self-linking numbers that arise
in Chern-Simons theory \cite{bs} which are regularization dependent.
Physically this is understood as an inducing of an intrinsic spin to the
$p-1$-branes by the doubling of field theoretical degrees of freedom in the
present case, which then cancels the statistical phases which appear because
the spin-statistics theorem holds in these special situations. This is in
contrast to the case of strings in three spatial dimensions, whereby the
spin-statistics theorem need not hold in certain cases \cite{bmosss}. The
generalizations of the standard Gauss linking integral are also derived.

\newsection{Global Aspects of Topological $BF$ Field Theory}

In this section we shall describe some general aspects of $BF$-type 
topological field theories. Starting with a quick review of the well-known
situation when the theory is defined using flat vector bundles, we then
describe the required modifications of the field theory when the associated
bundle lives in a non-trivial topological class.

\newsubsection{$BF$ Theory on Trivial Line Bundles}

Consider the field theory of a real-valued $p$-form field $B\in\Omega^p(\MDD)$
and a real-valued $d-p$-form field $A\in\Omega^{d-p}(\MDD)$ defined on the
spacetime manifold ${\cal M}_{d+1}$ of dimension $d+1$ with metric of Minkowski
signature. We assume in this subsection that these differential forms take
values in some flat vector bundle over $\MDD$. The $BF$-action is given by the
space-time integral of a $d+1$-form
\beq
S=\int\limits_{{\cal M}_{d+1}}{k\over2\pi}\,B\wedge dA
\label{BFdef}\eeq
where $k\in\IR$ is a coupling constant. This action is invariant under the
$U(1)$ gauge transform
\beq
A\rightarrow A+\chi
\label{Atransf}\eeq
where $\chi $ is a closed $d-p$-form, $d\chi=0$, and it transforms by a
surface term under
\beq
B\rightarrow B+\xi
\label{Btransf}\eeq
where $\xi$ is a closed $p$-form, $d\xi=0$. In the present field theory
without sources any closed forms are allowed in \eqn{Atransf} and
\eqn{Btransf}. However, when this topological field theory is coupled to
sources we also require gauge invariance of the holonomy operators
\beq
W[\Sigma_p]=\exp i\oint\limits_{\Sigma_p}B
{}~~~~~~,~~~~~~W[\Sigma_{d-p}]=\exp i\oint\limits_{\Sigma_{d-p}}A
\label{wilsonops}\eeq
This restricts the class of closed forms allowed in \eqn{Atransf} and
\eqn{Btransf} to those of integer-valued cohomology, so that
\beq
\oint\limits_{\Sigma_p}\xi=2\pi
n_p~~~~~~,~~~~~~\oint\limits_{\Sigma_{d-p}}\chi=2\pi n_{d-p}
\label{formrestr}\eeq
for some integers $n_p$ and $n_{d-p}$ and for any compact, closed orientable
submanifolds $\Sigma_p$ and $\Sigma_{d-p}$ of $\MDD$. In the following, we
shall assume this restricted gauge symmetry.

The partition function is given by the path integral
\beq
\langle1\rangle=\int\limits_\orbits D\mu(A,B)~\exp
i\int\limits_{\MDD}{k\over2\pi}\,B\wedge dA
\label{partfndef}\eeq
which is normalized by the volume of the gauge group. The functional measure
$D\mu(A,B)$ on the moduli space $\orbit$ of gauge orbits is obtained by the
standard gauge-fixing procedure to give\footnote{This is the conventional
parametrization for the partition function. In \cite{gk} an alternative
parametrization in terms of Hodge decompositions of the fields (see section
3.1) is implemented which avoids using the BRST gauge fixing procedure.}
\beq
D\mu(A,B)=DA~DB~\Delta_{\rm FP}[A]\,\Delta_{\rm FP}[B]\,\delta({\cal
F}[A])\,\delta({\cal G}[B])
\label{measure}\eeq
where $\Delta_{\rm FP}$ denotes the usual Faddeev-Popov determinant, and $\cal
F$ and $\cal G$ are gauge-fixing functions. This path integral is related to
the Ray-Singer analytic torsion which is a topological invariant of ${{\cal
M}_{d+1}}$ given by properties of the spectrum of the differential operators
$d$ and $\star\,d$ and the Laplacian ($\star$ denotes the Hodge duality
operator of $\MDD$). Here it is given explicitly by the ratio of determinants
\cite{schwarz,bt,gk,tqft}
\beq
\langle1\rangle=\prod_{k=0}^p{\det}^{\mu_k}_\perp
\,\Box_{p-k}~\prod_{l=0}^{d-p}{\det}_\perp^{\mu_l}
\,\Box_{d-p-l}~~~~~~,~~~~~~\mu_k\equiv(-1)^{k+1}\,\frac{2k+1}4
\label{raysinger}\eeq
where $\Box_k$ is the Laplacian acting on $\Omega^k(\MDD)$  and $\det_\perp$
denotes the regularized determinant with zero modes arising from gauge
invariance excluded.

Gauge and topologically invariant operators are given by $p$-cycle holonomies
of $B$ and $d-p$-cycle holonomies of $A$. The expectation value of the
operators \eqn{wilsonops} is given by the path integral with sources,
\beq
\frac{\left\langle
W[\Sigma_p]\,,\,W[\Sigma_{d-p}]\right\rangle}{\langle1\rangle}={\int_\orbits
D\mu(A,B)~\exp\left(i\int_{{\cal M}_{d+1}}{k\over2\pi}\,B\wedge
dA+i\oint_{\Sigma_{d-p}} A+ i\oint_{\Sigma_p}
B\right)\over\int_\orbits D\mu(A,B)~\exp i\int_{{\cal M}_{d+1}}
{k\over2\pi}\,B\wedge dA}
\label{expvalue}\eeq
This functional integral is independent of the metric of ${{\cal M}_{d+1}}$
and is formally a topological invariant. It is related to the topological
linking number of disjoint, closed contractible hypersurfaces $\Sigma_p $ and
$\Sigma_{d-p}$, which can be seen by explicitly performing the integral to
obtain \cite{bt,hs,tqft}
\beq
\frac{\left\langle
W[\Sigma_p]\,,\,W[\Sigma_{d-p}]\right\rangle}{\langle1\rangle}=\exp-\frac{2\pi
i}k\,L(\Sigma_p,\Sigma_{d-p})
\label{wilsonopsavg}\eeq
where
\beq
L(\Sigma_p,\Sigma_{d-p})=\int\limits_{{\cal
S}(\Sigma_p)}\Delta_{\Sigma_{d-p}}=(-1)^{(p-1)(d-p)}\int\limits_{{\cal
S}(\Sigma_{d-p})}\Delta_{\Sigma_p}
\label{linkingno}\eeq
is the standard expression for the signed linking number of two cycles. Here
${\cal S}(\Sigma_p)$ is a hypersurface spanned by the $p$-cycle $\Sigma_p$, and
$\Delta_{\Sigma_p}$ is the 
(singular) deRham current $d-p+1$-form \cite{botttu} which is the delta-function supported
Poincar\'e
dual to the embedding $X_p:\Sigma_p\to\MDD$. It is closed,
$d\Delta_{\Sigma_p}=0$, and locally it can be expressed as
\beq
\Delta_{\Sigma_p}(x)=\oint\limits_{\Sigma_p}\delta^{(p,d-p+1)}(X_p(\sigma),x)
=\star\oint\limits_{\Sigma_p}d\sigma(X_p)~\delta^{(d+1)}(X_p(\sigma),x)
\label{derhamcurrent}\eeq
where
\beq
d\sigma^{\mu_1\cdots\mu_p}(X_p)=\epsilon^{\alpha_1\cdots\alpha_p}\,
\prod_{k=1}^p\frac{\partial X_p^{\mu_k}(\sigma)}{\partial\sigma^{\alpha_k}}~d\sigma^k
\label{volelt}\eeq
is the induced volume element of $\Sigma_p$ in $\MDD$ and
$\delta^{(p,d-p+1)}(x,y)$ is the Dirac delta-function in the exterior algebra
$\Omega^p(\MDD(x))\otimes\Omega^{d-p+1}(\MDD(y))$, i.e.
\beq
\int\limits_{\MDD(y)}\delta^{(p,d-p+1)}(x,y)\wedge\alpha(y)
=\alpha(x)~~~~~~\forall\alpha(x)\in\Omega^p(\MDD(x))
\label{deltadef}\eeq

A more complete picture of this system is obtained by canonical quantization.
For this, we choose the spacetime to be the product manifold $\IR^1\times{{\cal
M}}_d$, where $\IR^1$ parametrizes the time $t$ and ${\cal M}_d$ is a compact,
path-connected, orientable $d$-dimensional manifold without
boundary.\footnote{In the following Greek indices $\mu=0,1,\dots,d$ will label
spacetime directions in $\MDD$ while Latin indices $i=1,\dots,d$ label spatial
directions in $\MD$. Furthermore, explicit metric factors required to make
quantities diffeomorphism invariant will be typically omitted.} The field $B$
may then be decomposed according to
\beq
B=B^0\wedge dt+\tilde B
\label{proddecomp}\eeq
where $B^0$ is the $p-1$-form on $\MD$ with local components $B^0_{i_1\cdots
i_{p-1}}=B_{i_1\cdots i_{p-1}0}$ and $\tilde B$ is the restriction of $B$ to
$\MD$ (and similarly for the other fields of the theory). The action is now
written as
\beq
S(\Sigma_p,\Sigma_{d-p})=\int\limits_{-\infty}^\infty dt~\int\limits_{{\cal
M}_d}\left({k\over2\pi}\,B\wedge
dA+Q_p\,B\wedge\Delta_{\Sigma_p}+Q_{d-p}\,A\wedge\Delta_{\Sigma_{d-p}}\right)
\label{canaction}\eeq
where $Q_p,Q_{d-p}\in\IR$ are worldvolume charges, $\Sigma_p$ and $\Sigma_{d-p}$
are disjoint hypersurfaces in $\MDD$, and we use the local
worldvolume reparametrization invariance to fix the gauge in which the temporal
embedding coordinate parametrizes the hypersurface $\Sigma_p$, i.e.
$X_p^0(\sigma^1,\dots,\sigma^p)=\sigma^1$. The temporal components of the
fields are Lagrange multipliers which enforce the local gauge constraints
\beq
{k\over2\pi}\,d\tilde
A+Q_p\,\tilde\Delta_{\Sigma_p}\approx0~~~~~~,~~~~~~(-1)^{p(d-p)}\,\frac
k{2\pi}\,d\tilde B+Q_{d-p}\,\tilde\Delta_{\Sigma_{d-p}}\approx0
\label{gaugeconstrs}\eeq
The remaining action is of first order in time derivatives and is therefore
already expressed in phase space with the spatial components of $A$ and $B$
being the canonically conjugate variables. The canonical quantum commutator is
\beq
\left[\tilde A_{i_1\cdots i_{d-p}}(x)\,,\,\tilde B_{j_1\cdots
j_{p}}(y)\right]={2\pi i\over k}\,\epsilon_{0i_1\cdots i_{d-p}j_1\cdots
j_{p}}\,\delta^{(d)}(x,y)
\label{cancomm}\eeq
Note that the factors of $\det g$, where $g$ is the metric of $\MD$, which
would make the delta-function on the right-hand side of \eqn{cancomm} generally
covariant cancel similar factors coming from the totally antisymmetric tensor
$\epsilon$. The canonical commutator is therefore independent of the metric of
${\cal M}_d$. The Hamiltonian in the temporal gauge $A^0=B^0=0$ is
\beq
H=-\int\limits_{{\cal M}_d}\left(Q_{d-p}\,\tilde
A\wedge\Delta^0_{\Sigma_{d-p}}+Q_p\,\tilde B\wedge\Delta^0_{\Sigma_p}\right)
\label{ham}\eeq

However, there is a technical problem with the way that we have set up the canonical
formalism for this topological field theory. The difficulty lies in the fact
that the local gauge constraints \eqn{gaugeconstrs}, which must be imposed as
physical state conditions in the quantum field theory, imply that the forms
$\tilde A$ and $\tilde B$ have non-vanishing flux around cycles of the manifold
$\MD$, i.e. generically we have $\oint_{\tilde\Sigma_{d-p+1}}d\tilde A\neq0$
and $\oint_{\tilde\Sigma_{p+1}}d\tilde B\neq0$. This implies that $A$ and $B$
cannot be considered as globally defined differential forms and must be defined
locally on patches covering the manifold. In turn, the definition of the
topological field theory must be appropriately modified. This will be the
subject of the next subsection. Here we note only one final aspect of this
model in the absence of sources ($Q_p=Q_{d-p}=0$). In that case the constraints
\eqn{gaugeconstrs} imply that $A$ and $B$ restricted to ${\cal M}_d$ are closed
forms. Their worldvolume  integrals are therefore topological invariants and
\eqn{cancomm} implies that they obey the operator algebra
\beq
\left[\oint\limits_{\tilde\Sigma_{d-p}}\tilde A\,,\,\oint\limits_{\tilde
\Sigma_p}\tilde
B\right]={2\pi i\over k}\,\nu[\tilde\Sigma_{d-p},\tilde\Sigma_p]
\label{opalg}\eeq
where
\beq
\nu[\tilde\Sigma_{d-p},\tilde
\Sigma_p]=\sum_{x\in\tilde\Sigma_p\cap\tilde\Sigma_{d-p}}{\rm sgn}(x)
\label{sgnint}\eeq
is the signed intersection number of the embedded hypersurfaces $\tilde\Sigma_{d-p}$
and $\tilde\Sigma_p$ on ${\cal M}_d$ taken over all intersections $x$ with
orientation ${\rm sgn}(x)=\pm1$ (Generically, in $d$ dimensions, a $p$-surface
and a $d-p$-surface intersect at discrete points). This number is a
topological invariant, so that if either $\tilde\Sigma_{d-p}$ or $\tilde
\Sigma_p$ is a
contractible hypersurface, then $\nu[\tilde\Sigma_{d-p},
\tilde\Sigma_p]$ vanishes and they
intersect an even number of times with cancelling orientations. Thus, in
\eqn{opalg}, the commutator is non-trivial only for those worldvolumes which
are non-trivial elements of the $p$-th and $d-p$-th homology groups $H_p(\MD)$
and $H_{d-p}(\MD)$. This property will be the crucial aspect of the topological
group representations that we shall find.

\newsubsection{$BF$ Theory on Non-trivial Line Bundles}

We shall now consider the case where the fields $A$ and $B$ are sections of some
non-trivial bundle over the spacetime manifold $\MDD$. Since these sections
have rank which is in general larger than 1, they are actually sections of a
higher-rank bundle \cite{brylinski} over $\MDD$, i.e. a fiber bundle whose
fibers are groupoids, rather than some Lie group. Such generalized
fiber bundles are known as gerbes. As pointed out in \cite{newbf}, in this case
the zero mode contribution to the path integral \eqn{partfndef} is modified and
the resulting topological invariant represents not just the Ray-Singer torsion
\eqn{raysinger}, but also the \v{C}ech cohomology of the underlying manifold.
Thus the canonical quantization of the $BF$ field theory in this case will
yield not only interesting quantum field theoretical representations of the
deRham complex of the spatial manifold $\MD$, but also of the more general
\v{C}ech complex which encodes the possibility of passage from local to global
data on $\MD$ and which classifies the topological line bundles over $\MD$.
Gerbes have also been used recently for some general analyses in quantum field 
theory \cite{cmm} and in the context of massive D-brane configurations
in Type II superstring theory \cite{kalk}.

We assume that the spatial manifold $\MD$ admits a finite open Leray cover
${\cal U}=\{U_a\}$. For each ordered collection
$(U_{a_0},U_{a_1},\dots,U_{a_q})$ of open sets of $\cal U$ with non-empty
intersection, we define the support
\beq
U_{a_0a_1\cdots a_q}=U_{a_0}\cap U_{a_1}\cap\cdots\cap U_{a_q}
\label{support}\eeq
along with a formal orientation defined by $U_{a_{\pi(0)}a_{\pi(1)}\cdots
a_{\pi(q)}}={\rm sgn}(\pi)\,U_{a_0a_1\cdots a_q}$ for $\pi\in S_{q+1}$. The
abelian group of all formal linear combinations with integer coefficients of
objects of the form \eqn{support} is the $q$-chain group $C_q({\cal U})$ of the
cover $\cal U$. Using a $\zed$-linear boundary operator $\partial$ defined on
$q$-chains by
\beq
\partial U_{a_0a_1\cdots a_q}=\sum_{k=0}^q(-1)^k\,U_{a_0a_1\cdots
a_{k-1}a_{k+1}\cdots a_q}
\label{bdryUdef}\eeq
one may define the $q$-th \v{C}ech homology group $H_q(\MD)$ which is
independent of the choice of Leray cover (If $\cal U$ is not a Leray cover,
then the homology must be defined by taking an inductive limit over all
possible open coverings of the space). 

The cover $\cal U$ naturally
defines a simplicial decomposition of the manifold $\MD$. Namely, to each
elementary $q$-chain \eqn{support} we can naturally associate a $q$-simplex
$\triangle_q$. In dimension $d$, we may label the simplices with a number of
indices which is dual to their dimension, by the inductive definition that a
$q$-simplex $\triangle_q^{(a_1\cdots a_{d-q})}$ is obtained as the intersection
of $d-q+1$ simplices of dimension $q+1$,
\bea
\triangle_q^{(a_1\cdots a_{d-q})}&=&\triangle_{q+1}^{(a_1\cdots
a_{d-q-1})}\cap\triangle_{q+1}^{(a_{d-q-1}a_{d-q}a_1\cdots
a_{d-q-3})}\cap\triangle_{q+1}^{(a_{d-q-3}a_{d-q-2}a_{d-q-1}a_{d-q}a_1\cdots
a_{d-q-5})}\nn\\& &\cap\cdots\cap\triangle_{q+1}^{(a_2\cdots a_{d-q})}
\label{simplexgen}\eea
along with the appropriate orientation induced by the supports of the cover
$\cal U$. With this convention the boundary operator acts on a $q$-simplex as
\beq
\partial\triangle_q^{(a_1\cdots
a_{d-q})}=\sum_a\sum_{k=1}^{d-q+1}(-1)^{k+1}\,\triangle_{q-1}^{(a_1\cdots
a_{k-1}aa_k\cdots a_{d-q})}
\label{bdrysimplexdef}\eeq
where the first sum in \eqn{bdrysimplexdef} runs through all $q-1$-simplices
with the appropriate index labellings. In this way the \v{C}ech homology of the
manifold $\MD$ coincides with its simplicial homology (This is also true if
$\MD$ doesn't admit a Leray cover).

On each open set $U_a$ of the covering $\cal U$ there is a $(d-p)$-form gauge
field $A^{(a)}$ and a $p$-form gauge field $B^{(a)}$, which are elements of the
0-cochain group $C^0({\cal U},\Omega^{d-p})$ with coefficients in the sheaf
$\Omega^{d-p}$ of real-valued differential $d-p$-forms on $\MD$ and of the
0-cochain group $C^0({\cal U},\Omega^p)$ with coefficients in the sheaf
$\Omega^p$ of $p$-forms, respectively. There are gauge transformations
$\Lambda_1^{(ab)}$ and $\Xi_1^{(ab)}$ defined on the non-empty intersection
$U_{ab}$ of any two open sets by
\beq
A^{(a)}-A^{(b)}=d\Lambda_1^{(ab)}~~~~~~,~~~~~~B^{(a)}-B^{(b)}=d\,\Xi_1^{(ab)}
\label{Uabfns}\eeq
with $\Lambda_1^{(ab)}=-\Lambda_1^{(ba)}$ and $\Xi_1^{(ab)}=-\Xi_1^{(ba)}$.
These local forms are elements of the 1-cochain groups $C^1({\cal
U},\Omega^{d-p-1})$ and $C^1({\cal U},\Omega^{p-1})$, respectively. Since each
$\Omega^q$ is a fine sheaf, the corresponding \v{C}ech cohomology is trivial,
and so there are secondary gauge transformations $\Lambda_2^{(abc)}\in C^2({\cal
U},\Omega^{d-p-2})$ and $\Xi_2^{(abc)}\in C^2({\cal U},\Omega^{p-2})$ defined
on non-empty triple intersections $U_{abc}$ by
\beq
\Lambda_1^{(ab)}+\Lambda_1^{(bc)}+\Lambda_1^{(ca)}=d\Lambda_2^{(abc)}~~~~~~,
~~~~~~\Xi_1^{(ab)}+\Xi_1^{(bc)}+\Xi_1^{(ca)}=d\,\Xi_2^{(abc)}
\label{Uabcfns}\eeq
This procedure can be iterated inductively to higher degree cochains over
higher degree chains. Namely, for each $1\leq q<d-p$ there are cochains
$\Lambda_q^{(a_0a_1\cdots a_q)}\in C^q({\cal U},\Omega^{d-p-q})$ and for each
$1\leq q<p$ we have $\Xi_q^{(a_0a_1\cdots a_q)}\in C^q({\cal U},\Omega^{p-q})$
which, on each non-trivial elementary $q+1$-chain $U_{a_0a_1\cdots a_{q+1}}$,
satisfy the overlap relations
\bd
\Lambda_q^{(a_0a_1\cdots a_q)}+\Lambda_q^{(a_qa_{q+1}a_0a_1\cdots
a_{q-2})}+\Lambda_q^{(a_{q-2}a_{q-1}a_qa_{q+1}a_0a_1\cdots a_{q-4})}
\ed
\beq
+\dots+\Lambda_q^{(a_1\cdots a_{q+1})}=d\Lambda_{q+1}^{(a_0a_1\cdots a_{q+1})}
\label{Uqfns}\eeq
\bd
\Xi_q^{(a_0a_1\cdots a_q)}+\Xi_q^{(a_qa_{q+1}a_0a_1\cdots
a_{q-2})}+\Xi_q^{(a_{q-2}a_{q-1}a_qa_{q+1}a_0a_1\cdots a_{q-4})}
\ed
\beq
+\dots+\Xi_q^{(a_1\cdots a_{q+1})}=d\,\Xi_{q+1}^{(a_0a_1\cdots a_{q+1})}
\label{Urfns}\eeq
Finally, on each $p$-chain $U_{a_0a_1\cdots a_p}$ and each $d-p$-chain
$U_{a_0a_1\cdots a_{d-p}}$ we have
\bd
\Lambda_{d-p}^{(a_0a_1\cdots
a_{d-p})}+\Lambda_{d-p}^{(a_{d-p}a_{d-p+1}a_0a_1\cdots
a_{d-p-2})}+\Lambda_{d-p}^{(a_{d-p-2}a_{d-p-1}a_{d-p}a_{d-p+1}a_0a_1\cdots
a_{d-p-4})}
\ed
\beq
+\dots+\Lambda_{d-p}^{(a_1\cdots a_{d-p+1})}=\lambda_{a_0a_1\cdots
a_{d-p+1}}\nn
\label{elldef}\eeq
\bd
\Xi_p^{(a_0a_1\cdots a_p)}+\Xi_p^{(a_pa_{p+1}a_0a_1\cdots
a_{p-2})}+\Xi_p^{(a_{p-2}a_{p-1}a_pa_{p+1}a_0a_1\cdots
a_{p-4})}
\ed
\beq
+\dots+\Xi_p^{(a_1\cdots a_{p+1})}=\xi_{a_0a_1\cdots a_{p+1}}
\label{cdef}\eeq
The locally constant functions $\lambda_{a_0a_1\cdots a_{d-p+1}}$ and
$\xi_{a_0a_1\cdots a_{p+1}}$ are, respectively, $d-p+1$-cocycles and
$p+1$-cocycles of the \v{C}ech cohomology groups $H_{\rm C}^{d-p+1}(\MD,\IR)$
and $H_{\rm C}^{p+1}(\MD,\IR)$ of the manifold $\MD$ with coefficients in the
constant sheaf $\IR$ (Again this is independent of the choice of Leray cover).
These \v{C}ech cohomology groups are naturally isomorphic to the corresponding
deRham cohomology groups, $H_{\rm C}^q(\MD,\IR)\cong H_{\rm D}^q(\MD)$ \cite{cechcs}.

The $\lambda$'s and $\xi$'s satisfy the usual cocycle relations
\bea
\lambda_{a_0a_1\cdots a_{d-p+1}}+\lambda_{a_{d-p+1}a_{d-p+2}a_0a_1\cdots
a_{d-p-1}}+\dots+\lambda_{a_1\cdots
a_{d-p+2}}&=&0\label{lambda0}\\\xi_{a_0a_1\cdots
a_{p+1}}+\xi_{a_{p+1}a_{p+2}a_0a_1\cdots a_{p-1}}+\dots+\xi_{a_1\cdots
a_{p+2}}&=&0
\label{xi0}\eea
on $U_{a_0a_1\cdots a_{d-p+2}}$ and $U_{a_0a_1\cdots a_{p+2}}$, respectively,
and they are completely antisymmetric in their indices. Given a $d-p+1$-cycle
$\tilde\Sigma_{d-p+1}$ and a $p+1$-cycle $\tilde\Sigma_{p+1}$ of $\MD$, the
corresponding fluxes of the gauge fields $A$ and $B$ can be represented in
terms of the \v{C}ech cohomology classes using the overlap relations
\eqn{Uabfns}--\eqn{cdef} and repeated application of Stokes' theorem to write
\bea
F_0(\tilde\Sigma_{d-p+1})&\equiv&\oint\limits_{\tilde\Sigma_{d-p+1}}dA
~=~{\sum}_{\triangle_0^{(a_0a_1\cdots a_{d-p+1})}(\tilde\Sigma_{d-p+1})}
~\lambda_{a_0a_1\cdots a_{d-p+1}}\label{Afluxrels}\\H_0(\tilde\Sigma_{p+1})
&\equiv&\oint\limits_{\tilde\Sigma_{p+1}}dB~=~{\sum}_{\triangle_0
^{(a_0a_1\cdots a_{p+1})}(\tilde\Sigma_{p+1})}~\xi_{a_0a_1\cdots a_{p+1}}
\label{Bfluxrels}\eea
where the sums are taken over all 0-simplices $\triangle_0^{(a_0a_1\cdots
a_{d-p+1})}(\tilde\Sigma_{d-p+1})$ and $\triangle_0^{(a_0a_1\cdots
a_{p+1})}(\tilde\Sigma_{p+1})$ with respect to the induced simplicial
decompositions of $\tilde\Sigma_{d-p+1}$ and $\tilde\Sigma_{p+1}$,
respectively, from the restrictions of the cover $\cal U$ to these submanifolds
of $\MD$ (via a refinement of $\cal U$ if necessary). Equations \eqn{Afluxrels}
and \eqn{Bfluxrels} demonstrate explicitly the relationship between the \v{C}ech
and deRham cohomologies.

Now we come to the appropriate extension of the action \eqn{canaction} for the
present situation. For this, we shall compactify the time direction on a
circle, so that $\MDD={\bf S}^1\times\MD$, and extend the Leray cover $\cal U$ with
its corresponding simplicial decomposition of $\MD$ trivially through the time
direction. This means that we shall consider only periodic motions on the space
$\MD$. Since the gauge fields $A$ and $B$ are only locally defined on $\MDD$,
the $BF$ action must be modified so that it is independent of the simplicial
decomposition (or covering) used to define the fields. Consider the term
\beq
\int\limits_{\MDD}B\wedge
dA=\sum_a\int\limits_{\triangle_{d+1}^{(a)}}B^{(a)}\wedge dA
\label{BFlocal}\eeq
where the sum runs through all $d+1$-simplices of $\MDD$. If we deform the
simplex $\triangle_{d+1}^{(a)}$, then the corresponding change of integrand in
\eqn{BFlocal} is $d(\Xi_1^{(ab)}\wedge dA)$, so that we must add the term
$-\sum_{a,b}\int_{\triangle_d^{(ab)}}\Xi_1^{(ab)}\wedge dA$ in order to cancel this
variation. In turn, we must cancel the dependence of this additional term on
deformations of the simplices $\triangle_d^{(ab)}$, and so on. Now we must
repeat this procedure for the deRham currents appearing in \eqn{canaction}.
Since $\Delta_{\Sigma_p}$ is a closed form representing the Poincar\'e
cohomology class of the cycle $\Sigma_p$, by Poincar\'e's lemma we have
\beq
\Delta_{\Sigma_p}=d\,\delta_{\Sigma_p}^{(a)}
\label{deltadef1}\eeq
on $U_a\in{\cal U}$, with $\delta_{\Sigma_p}^{(a)}\in C^0({\cal
U},\Omega^{d-p})$. Proceeding as before, we then obtain a set of cochains
${\cal X}_q^{(a_0a_1\cdots a_q)}\in C^q({\cal U},\Omega^{d-p-q})$ for $1\leq
q<d-p$ defined by
\beq
\delta_{\Sigma_p}^{(a)}-\delta_{\Sigma_p}^{(b)}=d{\cal X}_1^{(ab)}
\label{X1def}\eeq
\bd
{\cal X}_q^{(a_0a_1\cdots a_q)}+{\cal X}_q^{(a_qa_{q+1}a_0a_1\cdots
a_{q-2})}+{\cal X}_q^{(a_{q-2}a_{q-1}a_qa_{q+1}a_0a_1\cdots
a_{q-4})}
\ed
\beq
+\dots+{\cal X}_q^{(a_1\cdots a_{q+1})}=d{\cal
X}_{q+1}^{(a_0a_1\cdots a_{q+1})}
\label{DeltaU}\eeq
on $U_{ab}$ and $U_{a_0a_1\cdots a_{q+1}}$, respectively.

Using the method described above, we arrive at the consistent topological
extension of the source coupled $BF$ action \eqn{canaction},
\bea
S(\Sigma_p,\Sigma_{d-p})&=&\sum_a\int\limits_{\triangle_{d+1}^{(a)}}
\left[B^{(a)}\wedge\left(\frac k{2\pi}\,dA+Q_p\,\Delta_{\Sigma_p}\right)
+Q_{d-p}\,A^{(a)}\wedge\Delta_{\Sigma_{d-p}}\right]\nn\\& &
+\sum_{q=1}^p(-1)^q\sum_{a_0,a_1,\dots,a_{q+1}}\int
\limits_{\triangle_{d-q+1}^{(a_0a_1\cdots a_{q+1})}}
\Xi_q^{(a_0a_1\cdots a_q)}\wedge\left(\frac k{4\pi}\,dA
+Q_p\,\Delta_{\Sigma_p}\right)\nn\\& &+\sum_{q=1}^{d-p}(-1)^q
\sum_{a_0,a_1,\dots,a_q}\int\limits_{\triangle_{d-q+1}^{(a_0a_1\cdots a_q)}}
Q_{d-p}\,\Lambda_q^{(a_0a_1\cdots a_q)}\wedge\Delta_{\Sigma_{d-p}}\nn\\& &
-(-1)^p\sum_{a_0,a_1,\dots,a_{p+1}}\int\limits_{\triangle_{d-p}
^{(a_0a_1\cdots a_{p+1})}}\xi_{<a_0a_1\cdots a_{p+1}}
\left(\frac k{2\pi}\,A^{(a_{p+1})>}+Q_p\,\delta_{\Sigma_p}^{(a_{p+1})>}
\right)\nn\\& &-(-1)^{d-p}\sum_{a_0,a_1,\dots,a_{d-p+1}}\int
\limits_{\triangle_p^{(a_0a_1\cdots a_{d-p+1})}}Q_{d-p}\,
\lambda_{<a_0a_1\cdots a_{d-p+1}}\,\delta_{\Sigma_{d-p}}^{(a_{d-p+1})>}
\nn\\& &-\sum_{q=1}^{d-p-1}(-1)^{p+q}\sum_{a_0,a_1,\dots,a_{p+q+1}}
\int\limits_{\triangle_{d-p-q}^{(a_0a_1\cdots a_{p+q+1})}}
\xi_{<a_0a_1\cdots a_{p+1}}\nn\\& &\times\left(\frac k{2\pi}\,
\Lambda_q^{(a_{p+1}\cdots a_{p+q+1})>}+Q_p\,{\cal X}_q
^{(a_{p+1}\cdots a_{p+q+1})>}\right)-\sum_{q=1}^{p-1}(-1)^{d-p+q}
\nn\\& &\times\sum_{a_0,a_1,\dots,a_{d-p+q+1}}\int
\limits_{\triangle_{p-q}^{(a_0a_1\cdots a_{d-p+q+1})}}Q_{d-p}\,
\lambda_{<a_0a_1\cdots a_{d-p+1}}\,{\cal X}_q^{(a_{d-p+1}\cdots
a_{d-p+q+1})>}\nn\\& &-(-1)^d\sum_{a_0,a_1,\dots,a_{d+1}}
\left[\xi_{<a_0a_1\cdots a_{p+1}}\left(\frac k{2\pi}\,
\Lambda_{d-p}^{(a_{p+1}\cdots a_{d+1})>}+Q_p\,{\cal X}_p^{(a_{p+1}\cdots
a_{d+1})>}\right)\right.\nn\\& &\biggl.+\,Q_{d-p}\,
\lambda_{<a_0a_1\cdots a_{d-p+1}}\,{\cal X}_p^{(a_{d-p+1}\cdots
a_{d+1})>}\biggr]\left(\triangle_0^{(a_0a_1\cdots a_{d+1})}\right)
\label{actionext}\eea
where the sums all run over the simplicial decomposition of the spacetime
manifold $\MDD$. The brackets $<\cdots>$ in \eqn{actionext} act on the indices
of a product of cochains by putting them in increasing order with repeated
indices matched according to their position in the sequence (not their value)
and then multiplying the cochain product by the parity of the rearranging
permutation, i.e. for $\pi\in S_d$ we have
\beq
X^{<(\pi_1\cdots\pi_p)}\,Y^{(\pi_p\cdots\pi_d)>}={\rm sgn}(\pi)\,X^{(1\cdots
p)}\,Y^{(p\cdots d)}
\label{bracketdef}\eeq
It is straightforward to verify that the complicated expression \eqn{actionext}
ensures that the total action is independent of the simplicial decomposition of
$\MDD$ used to evaluate each of the integrals. In this way, we arrive at a
topologically invariant action for the $BF$ field theory with gauge fields that
are sections of a non-trivial $U(1)$ bundle over the manifold.

The action \eqn{actionext} as it stands is difficult to deal with, especially
for the quantum field theory. However, it can be simplified by noticing that
the gauge field $A$ may be decomposed in terms of an arbitrary globally-defined
differential form ${\cal A}\in\Omega^{d-p}(\MDD)$ and a singular form ${\bb A}$
which is
an explicit representative of the topological bundle of $A$
(and similarly for $B$). This latter degree
of freedom may be constructed as follows. Let $N_p$ be the rank of the 
singular homology
groups $H_p(\MD)$ and $H_{d-p}(\MD)$, and let $\tilde\Sigma_p^{(k)}$ and
$\tilde\Sigma_{d-p}^{(k)}$ be sets of corresponding generators. The associated
intersection matrix of $\MD$ is
\beq
I^{(p)kl}=\oint\limits_{\tilde\Sigma_{d-p}^{(l)}(x)}\oint\limits_{\tilde
\Sigma_p^{(k)}(y)}\delta^{(d-p,p)}(x,y)
\label{intpmatrix}\eeq
Then the gauge field $A$ may be written as
\beq
A=\sum_{k=1}^{N_{p-1}}{\bb A}_k+{\cal A}
\label{Asingdecomp}\eeq
where the singular forms ${\bb A}_k$ are defined by
\beq
d{\bb
A}_k=(-1)^{p(d-p)}\sum_{l=1}^{N_{p-1}}I_{lk}^{(p-1)}\,
F_0(\tilde\Sigma_{d-p+1}^{(l)})\,\tilde\Delta_{\tilde\Sigma_{p-1}^{(k)}}
\label{bbAdef}\eeq
and they ensure that $A$ has the correct periods \eqn{Afluxrels} around cycles
of $\MD$. Here $I_{lk}^{(p)}$ is the matrix inverse of the intersection
matrix \eqn{intpmatrix}, and $\tilde\Delta$ denotes the deRham current which 
is Poincar\'e dual to a cycle in the {\it spatial} manifold $\MD$ (for
ease of notation the tildes are suppressed on Dirac delta-functions over
$\MD$, as in \eqn{intpmatrix}). This means that we have chosen to
localize the flux of the gauge fields over submanifolds of $\MD$, rather
than the full spacetime manifold $\MDD$. This choice is a necessary requirement
in the canonical formalism, and is possible due to the topological 
triviality of the time direction of $\MDD$.
Similarly, the gauge field $B$ can be written as
\beq
B=\sum_{k=1}^{N_{p+1}}{\bb B}_k+{\cal B}
\label{Bsingdecomp}\eeq
where ${\cal B}\in\Omega^p(\MDD)$ is a globally-defined differential form and
${\bb B}_k$ are singular forms defined by
\beq
d\,{\bb
B}_k=\sum_{l=1}^{N_{p+1}}I_{lk}^{(p+1)}\,H_0(\tilde\Sigma_{p+1}^{(l)})
\,\tilde\Delta_{\tilde\Sigma_{d-p-1}^{(k)}}
\label{bbBdef}\eeq
which ensure that $B$ has the correct periods \eqn{Bfluxrels}. Finally, the
deRham currents are written as
\beq
\Delta_{\Sigma_p}=\sum_{k,l=1}^{N_{p-1}}I_{lk}^{(p-1)}\,
\nu[\tilde\Sigma_{d-p+1}^{(l)},\partial\Sigma_p(0)]\,
\tilde\Delta_{\tilde\Sigma_{p-1}^{(k)}}+d\,\delta_{\Sigma_p}
\label{Dsingdecomp}\eeq
with $\delta_{\Sigma_p}\in\Omega^{d-p}(\MDD)$. Here $\Sigma_p(t)$
represents the embedded hypersurface $X_p(\Sigma_p)\subset\MDD$ projected
onto $\MD$ with boundary the $p-1$-brane $X_p(t,\sigma^2,\dots,\sigma^p)$
at time $t$, and we have localized the period integrals of the deRham 
currents onto a fixed patch of $\MD$ at $t=0$ (as with the fluxes in
\eqn{bbAdef} and \eqn{bbBdef}).

Substituting the decompositions \eqn{Asingdecomp}--\eqn{Dsingdecomp} into the
action \eqn{canaction} and integrating by parts we have
\bea
S(\Sigma_p,\Sigma_{d-p})&=&\sum_{m=1}^{N_{p-1}}\sum_{l=1}^{N_{p+1}}\oint
dt~\int\limits_{\MD}\left(\frac k{2\pi}\,{\bb B}_l\wedge d{\bb A}_m\right.
\nn\\& &+\,Q_p\sum
_{n=1}^{N_{p-1}}I_{nm}^{(p-1)}\,\nu[\tilde\Sigma_{d-p+1}^{(n)},\partial
\Sigma_p(0)]\,{\bb
B}_l\wedge\tilde\Delta_{\tilde
\Sigma_{p-1}^{(m)}}\nn\\& &\left.+\,Q_{d-p}\sum_{n=1}^{N_{p+1}}
I_{nl}^{(p+1)}\,\nu[\tilde\Sigma_{p+1}^{(n)},\partial\Sigma_{d-p}(0)]\,{\bb
A}_m\wedge\tilde\Delta_{\tilde\Sigma_{d-p-1}^{(l)}}\right)\nn\\& &+\oint
dt~\int\limits_{\MD}\left[\sum_{m=1}^{N_{p-1}}\left(\frac k{2\pi}\,{\cal
B}+(-1)^{p(d-p)}\,Q_{d-p}\,\delta_{\Sigma_{d-p}}\right)\wedge d{\bb
A}_m\right.\nn\\& &+\,(-1)^{p(d-p)}\sum_{m=1}^{N_{p+1}}\left(\frac
k{2\pi}\,{\cal A}+Q_p\,\delta_{\Sigma_p}\right)\wedge d\,{\bb B}_m\nn\\&
&+\left.\,\frac k{2\pi}\,{\cal B}\wedge d{\cal A}+Q_{d-p}\,{\cal
A}\wedge\Delta_{\Sigma_{d-p}}+Q_p\,{\cal B}\wedge\Delta_{\Sigma_p}\right]
\label{actiondecomp}\eea
The first set of integrals in \eqn{actiondecomp} are defined using the
topological extension \eqn{actionext}. Remembering that the simplicial
decomposition of $\MD$ is extended trivially through the periodic time
direction of $\MDD$, using the decompositions
\eqn{Asingdecomp}--\eqn{Dsingdecomp} and some calculation we find that the only
non-vanishing contributions are
\bea
S_{\rm
sing}&=&(-1)^d\sum_{k=1}^{N_{p-1}}\sum_{l=1}^{N_{p+1}}\left(Q_{d-p}
\sum_{m=1}^{N_{p+1}}I_{ml}^{(p+1)}\,\nu[\tilde\Sigma_{p+1}^{(l)},
\partial\Sigma_{d-p}(0)]\right.\nn\\& &\times\sum_{a_0,a_1,\dots,a_{d-p+1}}
\int\limits_{\triangle_p^{(a_0a_1\cdots a_{d-p+1})}}
\lambda^{(k)}_{<a_0a_1\cdots a_{d-p+1}}\,\delta_{\tilde
\Sigma_{d-p-1}^{(l)}}^{(a_{d-p+1})>}\nn\\& &
+\,Q_p\sum_{m=1}^{N_{p-1}}I_{mk}^{(p-1)}\,\nu[\tilde\Sigma_{d-p+1}^{(m)},
\partial\Sigma_p(0)]\nn\\& &\left.\times\sum_{a_0,a_1,\dots,a_{p+1}}\int
\limits_{\triangle_{d-p}^{(a_0a_1\cdots a_{p+1})}}
\xi^{(l)}_{<a_0a_1\cdots a_{p+1}}\,\delta_{\tilde
\Sigma_{p-1}^{(k)}}^{(a_{p+1})>}\right)
\label{Ssing}\eea
The contribution \eqn{Ssing} is not a topological invariant because it now
changes under deformations of the simplices of $\MD$. Using the period
relations \eqn{Afluxrels} and \eqn{Bfluxrels}, we see that 
the two sets of integrals in
\eqn{Ssing} are defined modulo terms of the form $Q_{d-p}F_0$ and $Q_pH_0$ (for
periodic motions). 
In the quantum field theory, such terms would then
appear as phases $\e^{iS_{\rm sing}}$ and thus the ambiguity can be removed by
a flux relation among the gauge fields. The required consistency condition
ensuring topological invariance of the quantum field theory is thus
\bea
1&=&\prod_{l=1}^{N_{p-1}}\prod_{k=1}^{N_{p+1}}\exp i\,Q_{d-p}F_0(\tilde
\Sigma_{d-p+1}^{(l)})\sum_{m=1}^{N_{p+1}}
I_{mk}^{(p+1)}\,\nu[\tilde\Sigma_{p+1}^{(m)},\partial\Sigma_{d-p}(0)]
\nn\\& &\times\prod_{l=1}^{N_{p-1}}\prod_{k=1}^{N_{p+1}}\exp
i\,Q_pH_0(\tilde\Sigma_{p+1}^{(k)})\sum_{m=1}^{N_{p-1}}I_{mk}^{(p-1)}
\,\nu[\tilde\Sigma_{d-p+1}^{(m)},\partial\Sigma_p(0)]
\label{conscondn}\eea
In the following sections we will see that this condition does indeed lead to a
sensible Hilbert space representation. We shall see later on that it implies
some noteworthy features of the motion group on topologically non-trivial
spaces.

Next, let us consider the integrals
\beq
\oint dt~\int\limits_{\MD}\delta_{\Sigma_p}\wedge d\,{\bb
B}_k=\sum_{l=1}^{N_{p+1}}I_{lk}^{(p+1)}
H_0(\tilde\Sigma_{p+1}^{(k)})~\oint\limits_{\tilde\Sigma_{d-p-1}^{(k)}}
\delta_{\Sigma_p}^0\wedge dt
\label{deltaint0}\eeq
The integration in \eqn{deltaint0} can be set to 0 via a judicious choice of
decomposition of the fixed deRham currents. This will be done in the next
section. The essential feature is that
the deRham currents are decomposed in \eqn{Dsingdecomp} so that the
only non-vanishing period integrals come from cycles which lie entirely in the
spatial manifold $\MD$. A similar calculation shows that the fourth and fifth
lines of \eqn{actiondecomp} can be taken to be 0 in the temporal gauge. The final
result is the action
\beq
S(\Sigma_p,\Sigma_{d-p})=S_{\rm sing}+\oint dt~\int\limits_{\MD}\left(\frac
k{2\pi}\,{\cal B}\wedge d{\cal A}+Q_{d-p}\,{\cal
A}\wedge\Delta_{\Sigma_{d-p}}+Q_p\,{\cal B}\wedge\Delta_{\Sigma_p}\right)
\label{Sfinal}\eeq
This form of the action, along with the constraint \eqn{conscondn}, will
be used to construct the canonical quantum field theory in the following sections.

\newsection{Canonical Quantization}

Having introduced the required modifications of the topological field theory
that are required over non-trivial bundles, we shall now proceed to study the
structure of the phase space of this system which will be used in the next section
to construct the wavefunctions of the canonical quantum field theory.

\newsubsection{Hodge Decompositions}

To deal with the quantum field theory associated with the action \eqn{Sfinal},
it will be convenient to exploit the fact that $\cal A$ and $\cal B$ are
globally defined differential forms on $\MD$ and therefore admit Hodge
decompositions \cite{botttu}.
In this way we may write the field degrees of freedom in terms
of their local exact and co-exact components, and their global components which
take into account the topological degrees of freedom. For this, we
consider the intersection matrix $I^{(p)kl}$ of
$\MD$ whose matrix inverse can be defined by the bilinear form
\beq
I_{kl}^{(p)}=\int\limits_{\MD}\alpha_l^{(p)}\wedge\beta_k^{(p)}
\label{Ikldef}\eeq
where $\{\alpha_l^{(p)}\}_{l=1}^{N_p}$ and $\{\beta_k^{(p)}
\}_{k=1}^{N_p}$ are bases
of generators of $H_{\rm D}^{d-p}(\MD)$ and $H_{\rm D}^p(\MD)$, respectively,
which are orthonormal in the inner product
\beq
\int\limits_{\MD}\alpha^{(p)}_l\wedge*\alpha^{(p)}_k=
\int\limits_{\MD}\beta^{(p)}_l\wedge*\beta^{(p)}_k=\delta_{lk}
\label{abortho}\eeq
Here $*$ is the Hodge duality operator on $\MD$ with respect to the restriction
of the spacetime metric of $\MDD$. With this definition, the harmonic forms
$\alpha^{(p)}_l$ and $\beta^{(p)}_k$ are the (non-singular) 
Poincar\'e duals of the corresponding homology
generators $\tilde\Sigma_{d-p}^{(l)}$ and $\tilde
\Sigma_p^{(k)}$ of the free parts of the
singular homology groups $H_{d-p}(\MD)$ and $H_p(\MD)$, respectively (the torsion
components of the homology will play no role in what follows). The
natural bilinear pairing on deRham cohomology between any closed $d-p$-form
$\alpha$ and any closed $p$-form $\beta$ may then be written as
\beq
\int\limits_{\MD}\alpha\wedge\beta=\sum_{k,l=1}^{N_p}\left(\oint\limits_{
\tilde\Sigma_{d-p}^{(l)}}\alpha\right)I_{lk}^{(p)}\left(\oint\limits_{
\tilde\Sigma_p^{(k)}}\beta\right)
\label{pairing1}\eeq
We shall denote
by $\nabla_p^2=*d*d$ the Laplacian operator acting on co-closed $p$-forms in
$\Omega^p(\MD)$.

The field $\tilde{\cal A}$ may then be expressed in terms of its Hodge
decomposition over $\MD$ as
\beq
\tilde{\cal A}=d\theta+*dP_K+\sum_{l=1}^{N_p}a^l(t)\,\alpha^{(p)}_l
\label{Adecomp}\eeq
where $\theta\in\Omega^{d-p-1}(\MD)$ and $P_K\in\Omega^{p-1}(\MD)$ with
\bea
& &\nabla_{d-p-1}^2\theta=*d*\tilde{\cal
A}~~~~~~,~~~~~~\nabla_{p-1}^2P_K=*d\tilde{\cal A}\nonumber\\&
&a^l(t)=\sum_{k=1}^{N_p}I^{(p)kl}\int\limits_{\MD}\tilde{\cal
A}\wedge\beta_k^{(p)}=\oint\limits_{\tilde\Sigma_{d-p}^{(l)}}\tilde{\cal A}
\label{Acomps}\eea
Since, by assumption, $\tilde{\cal A}$ is a globally defined differential form
on $\MD$, the form $*d\tilde{\cal A}$ contains no zero modes (harmonic forms)
of the Laplacian operator $\nabla_{p-1}^2$. Moreover, the local and global
parts of the gauge transformations \eqn{Atransf} may be expressed in terms of
the above degrees of freedom as
\beq
\theta\to\theta+\chi'~~~~~~,~~~~~~a^l\to a^l+2\pi n_{d-p}^l
\label{Acomptransf}\eeq
where $d\chi'$ is the local exact part of the closed $d-p$-form $\chi$ and
$n_{d-p}^l$
labels the winding numbers of the gauge field $\tilde{\cal A}$ around the
Poincar\'e dual homology basis element $\tilde\Sigma_{d-p}^{(l)}$. Using the
time-independent gauge transformations \eqn{Acomptransf}, we can remove the
Laplacian zero modes of the form $*d*\tilde{\cal A}$. Similarly, the Hodge
decomposition of the field $\tilde{\cal B}$ over $\MD$ is
\beq
\tilde{\cal B}=*dP_\theta+dK+\sum_{l=1}^{N_p}b^l(t)\,\beta^{(p)}_l
\label{Bdecomp}\eeq
where $P_\theta\in\Omega^{d-p-1}(\MD)$ and $K\in\Omega^{p-1}(\MD)$ with
\bea
& &\nabla_{d-p-1}^2P_\theta=*d\tilde{\cal
B}~~~~~~,~~~~~~\nabla_{p-1}^2K=*d*\tilde{\cal B}\nonumber\\&
&b^l(t)=\sum_{k=1}^{N_p}I^{(p)kl}\int\limits_{\MD}\tilde{\cal
B}\wedge\alpha_k^{(p)}=\oint\limits_{\tilde\Sigma_{p}^{(l)}}\tilde{\cal B}
\label{Bcomps}\eea
and the gauge transformations \eqn{Btransf} may be written as
\beq
K\to K+\xi'~~~~~~,~~~~~~b^l\to b^l+2\pi n_p^l
\label{Bcomptransf}\eeq
where $d\xi'$ is the local exact part of the closed 
$p$-form $\xi$. It follows that the
harmonic modes of the differential forms $*d\tilde{\cal B}$ and $*d*\tilde{\cal
B}$ may be set to 0.

It will prove convenient to use a holomorphic polarization for the harmonic
degrees of freedom of the gauge fields. For this, we consider the
$2N_p$-dimensional symplectic vector space
\beq
{\cal P}=H_{\rm D}^p(\MD)\oplus H_{\rm D}^{d-p}(\MD)
\label{calPdef}\eeq
which, according to the gauge constraints \eqn{gaugeconstrs}, is the
reduced classical
phase space of the source-free $BF$ field theory and is spanned by the
topological degrees of freedom $a^l$ and $b^k$ of the gauge fields. On this
finite dimensional vector space we may introduce a complex structure which is
parametrized by an $N_p\times N_p$ symmetric complex-valued matrix
$\tau$ such that $-\tau$ lives in the Siegal upper half-plane. Its imaginary
part defines a metric
\beq
G^{lk}=-2\sum_{m,n=1}^{N_p}I^{(p)ml}\,\left({\rm
Im}\,\tau_{mn}\right)\,I^{(p)nk}
\label{Gdef}\eeq
on the topological phase space $\cal P$. Note that the topological invariance
property of the $BF$ field theory implies that all observables will be
independent of the phase space complex structure. The desired holomorphic
polarization is then defined by the complex variables
\beq
\gamma^l=a^l+\sum_{k,m=1}^{N_p}
I^{(p)ml}\tau_{mk}\,b^k~~~~~~,~~~~~~\overline{\gamma}^l=a^l+\sum_{k,m=1}^
{N_p}I^{(p)ml}\overline{\tau}_{mk}\,b^k
\label{gammadef}\eeq
in terms of which the large gauge transformations
take the form
\beq
\gamma^l\to\gamma^l+2\pi\left(n_{d-p}^l+\sum_{k,m=1}^{N_p}
I^{(p)ml}\tau_{mk}\,n_p^k\right)~~~~~~,~~~~~~\overline{\gamma}^l\to
\overline{\gamma}^l+2\pi\left(n_{d-p}^l+\sum_{k,m=1}^{N_p}I^{(p)ml}
\overline{\tau}_{mk}\,n_p^k\right)
\label{largegaugegamma}\eeq

Now we come to the Hodge decompositions for the non-singular parts of the
deRham currents \eqn{Dsingdecomp}. We have
\beq
\star d\,\delta_{\Sigma_p}=\bar \delta_{\Sigma_p}^0\wedge
dt+\tilde\delta_{\Sigma_p}
\label{starD}\eeq
where the $p$-form $\tilde\delta_{\Sigma_p}$ may be decomposed as
\beq
\tilde\delta_{\Sigma_p}=d\omega_p+*d\rho_p+\sum_{k,l=1}^
{N_p}\Sigma_l^{(p)}(t)\,I^{(p)kl}*\alpha_k^{(p)}
\label{Ddecomp}\eeq
with $\omega_p\in\Omega^{p-1}(\MD)$ and $\rho_p\in\Omega^{d-p-1}(\MD)$. From
the source continuity equation $d^2\delta_{\Sigma_p}=0$ we have
\beq
\frac\partial{\partial
t}\,\bar\delta_{\Sigma_p}^0=-*d*\tilde\delta_{\Sigma_p}=-\nabla_{p-1}^2\omega_p
\label{conteq}\eeq
and from the definitions we find
\bea
*d\tilde\delta_{\Sigma_p}&=&\nabla_{d-p-1}^2\rho_p\label{Drhocomp}\\
\Sigma_l^{(p)}(t)&=&\sum_{k=1}^{N_p}I^{(p)}_{lk}\int\limits_{\MD}\tilde
\delta_{\Sigma_p}\wedge\alpha^{(p)}_k~=~\frac d{dt}\int\limits_{\Sigma_p(t)}
\beta^{(p)}_l
\label{Dcomps}\eea
In arriving at the second
equality in \eqn{Dcomps} we have used Poincar\'e-Hodge
duality and the local form \eqn{derhamcurrent} of the deRham current. Again the
globally defined differential form $*d\tilde\delta_{\Sigma_p}$ contains no zero
modes of the Laplacian operator $\nabla_{d-p-1}^2$, and the continuity equation
\eqn{conteq} implies the current ``gauge symmetry''
\beq
\rho_p\to\rho_p+\tilde\Lambda
\label{currentsym}\eeq
with $\tilde\Lambda$ an arbitrary $d-p-1$-form, which allows one to remove the
harmonic components of $*d*\tilde\delta_{\Sigma_p}$.

Finally, we shall write the Hodge decompositions \eqn{Ddecomp} of the deRham
currents in terms of a generalized eigenfunction expansion on $\MD$. For this,
we introduce, for each $p$, a basis of co-exact $p$-forms
$\psi_{\lambda_p}^{(p)}$ which constitute the complete system of eigenstates of
the Laplacian operator $\nabla_p^2$ with eigenvalues $\lambda_p^2\geq0$:
\beq
*d*\psi_{\lambda_p}^{(p)}=0~~~~~~,~~~~~~\nabla_p^2\psi_{\lambda_p}^{(p)}
=*d*d\psi_{\lambda_p}^{(p)}=\lambda_p^2\,\psi_{\lambda_p}^{(p)}
\label{psilambdapdef}\eeq
and which are orthonormal:
\beq
\int\limits_{\MD}\psi_{\lambda_p}^{(p)}\wedge*\psi_{\lambda_p'}^{(p)}
=\delta_{\lambda_p,\lambda_p'}
\label{orthopsi}\eeq
Because of Hodge duality, we may identify $\psi_{\lambda_p}^{(p)}=*d\psi_{\lambda
_{d-p-1}}^{(d-p-1)}$, and so it suffices to consider only $[\frac d2]$ of these
$p$-form eigenfunctions. Note that the zero modes of $\nabla_p^2$ are just the
harmonic $p$-forms, $\psi_0^{(p)}=\{\beta^{(p)}_k\}_{k=1}^{N_p}$.

These eigenstates are particularly useful for expanding the Dirac
delta-functions which act on the exterior algebras of $\MD$ in terms of
completeness relations. For example, we can readily write down the following
distribution-valued Hodge decompositions over the appropriate exterior
algebras:
\bea
\delta^{(d)}(x,y)&=&\sum_{\lambda_0}\psi_{\lambda_0}^{(0)}(x)\,
\psi_{\lambda_0}^{(0)}(y)\nonumber\\\delta^{(d-p,p)}(x,y)&=&
-\sum_{\lambda_{d-p-1}\neq0}\frac1{\lambda_{d-p-1}^2}\,
d\psi_{\lambda_{d-p-1}}^{(d-p-1)}(x)\otimes*d\psi_{\lambda_{d-p-1}}^{(d-p-1)}
(y)\nonumber\\& &+\,\sum_{\lambda_{d-p}\neq0}\psi_{\lambda_{d-p}}^{(d-p)}(x)
\otimes*\psi_{\lambda_{d-p}}^{(d-p)}(y)+\sum_{k,l=1}^{N_p}\alpha^{(p)}_l(x)
\otimes I^{(p)kl}\,\beta^{(p)}_k(y)\nn\\& &
\label{diracdecomps}\eea
Then, by equating the exact and co-exact parts of $\delta^{(d-p,p)}(x,y)$ in
\eqn{diracdecomps} with those of $\star d\delta_{\Sigma_p}$ in \eqn{Ddecomp},
we arrive the following generalized eigenfunction expansions for the components
of the deRham currents:
\bea
\bar\delta_{\Sigma_p}^0&=&\sum_{\lambda_{p-1}}\psi_{\lambda_{p-1}}^{(p-1)}
\oint\limits_{\partial\Sigma_p(t)}\psi_{\lambda_{p-1}}^{(p-1)}\nonumber\\
\rho_p&=&-\sum_{\lambda_{d-p-1}\neq0}\frac1{\lambda_{d-p-1}^2}\,
\psi_{\lambda_{d-p-1}}^{(d-p-1)}\,\frac d{dt}\int\limits_{\Sigma_p(t)}
*d\psi_{\lambda_{d-p-1}}^{(d-p-1)}\nonumber\\\omega_p&=&
-\sum_{\lambda_{p-1}\neq0}\frac1{\lambda_{p-1}^2}\,\psi_{\lambda_{p-1}}^{(p-1)}
\,\frac d{dt}\oint\limits_{\partial\Sigma_p(t)}\psi_{\lambda_{p-1}}^{(p-1)}
\label{Deigen}\eea
where we have used \eqn{conteq}, \eqn{Dcomps}, \eqn{derhamcurrent}, and Hodge
duality. Note that with the decomposition of $\bar\delta_{\Sigma_p}^0$ in
\eqn{Deigen}, the integral \eqn{deltaint0} vanishes because it yields the
intersection number of the projected hypersurface $\Sigma_p(t)$ with the time
direction of $\MDD$, which is 0.

\newsubsection{Canonical Structure}

We shall now proceed to describe the canonical quantization of the field
theory. First we examine the local gauge constraints \eqn{gaugeconstrs}.
Upon substitution of the decompositions \eqn{Asingdecomp}, \eqn{bbAdef} and
\eqn{Dsingdecomp}, we may integrate the multi-valued part of \eqn{gaugeconstrs}
over a suitable cycle of $\MD$ and obtain the relation
\beq
F_0(\tilde\Sigma_{d-p+1}^{(l)})=-\frac{2\pi}k\,Q_p\,\nu[\tilde\Sigma_{d-p+1}
^{(l)},\partial\Sigma_p(0)]
\label{FQrel}\eeq
Note that this relation is a strong equality since neither the flux $F_0$ nor
the charge $Q_p$ will be a dynamical degree of freedom. The remaining
non-singular part of \eqn{gaugeconstrs} can be expressed in terms of the
decompositions \eqn{Acomps} and \eqn{starD} to give the weak equality
\beq
(-1)^{(d-1)(p-1)}\,\frac
k{2\pi}\,\nabla_{p-1}^2P_K+Q_p\,\bar\delta_{\Sigma_p}^0\approx0
\label{Aweakeq}\eeq
Similarly, from the local gauge constraints associated with the $\tilde B$
field in \eqn{gaugeconstrs} we obtain, using \eqn{Bsingdecomp} and
\eqn{bbBdef}, the strong equality
\beq
H_0(\tilde\Sigma_{p+1}^{(l)})=-(-1)^{p(d-p)}\,\frac{2\pi}k\,Q_{d-p}\,
\nu[\tilde\Sigma_{p+1}^{(l)},\partial\Sigma_{d-p}(0)]
\label{HQrel}\eeq
and from \eqn{Bcomps} the weak equality
\beq
-(-1)^d\,\frac
k{2\pi}\,\nabla_{d-p-1}^2P_\theta+Q_{d-p}\,
\bar\delta^0_{\Sigma_{d-p}}\approx0
\label{Bweakeq}\eeq
Using the strong equalities \eqn{FQrel} and \eqn{HQrel}, we can now rewrite the
consistency condition \eqn{conscondn} in the form
\beq
1=\prod_{r,l=1}^{N_{p-1}}\,\prod_{m,n=1}^{N_{p+1}}\exp-\frac{2\pi 
i}k\,Q_p\,Q_{d-p}\,\nu[\tilde\Sigma_{p+1}^{(m)},\partial\Sigma_{d-p}(0)]
\,\nu[\tilde\Sigma_{d-p+1}^{(l)},\partial\Sigma_p(0)]
\left(I_{mn}^{(p+1)}+I_{rl}^{(p-1)}\right)
\label{topquant}\eeq
The constraint \eqn{topquant} is a topological quantization condition determined
by the intersection numbers of the homology cycles
of $\MD$. It is the appropriate generalization to the present case of the usual
Dirac charge quantization condition.
The global gauge constraints of the $BF$ field theory will be
described in the next section.

Having described the gauge constraints of the model, we may now write down the
canonical quantum commutators. For this, we need to examine the source-free $BF$
action in \eqn{Sfinal}. Using the Hodge decompositions of the previous subsection
we may write down explicitly the remaining $BF$ action without the gauge
constraints (i.e. in the temporal gauge ${\cal A}^0={\cal B}^0=0$). After some
algebra and integrations by parts over $\MD$ it is straightforward to arrive at
\beq
S=\oint dt~\frac
k{2\pi}\left[\int\limits_{\MD}\left(*\dot{\theta}\wedge\nabla_{d-p-1}^2P_\theta
+\nabla_{p-1}^2K\wedge*\dot P_K\right)+(-1)^{p(d-p)}\,\frac i2
\sum_{k,l=1}^{N_p}G_{kl}\left(\overline{\gamma}^k\dot{\gamma}^l-
\gamma^k\dot{\overline{\gamma}}^l\right)\right]
\label{Srem}\eeq
where $G_{kl}$ is the matrix inverse of the topological phase space metric
\eqn{Gdef}. From the action \eqn{Srem} we can immediately read off the
non-vanishing canonical Poisson brackets
\bea
\Bigl\{\theta(x)\,{\buildrel\otimes\over,}\,P_\theta(y)\Bigr\}
&=&-\frac{2\pi}k\,\frac1{(\nabla_{d-p-1}^2)^\perp}\,\Pi^{(d-p-1)}\,
\delta^{(d)}(x,y)\nonumber\\\Bigl\{K(x)\,{\buildrel\otimes\over,}\,P_K(y)\Bigr\}
&=&-\frac{2\pi}k\,\frac1{(\nabla_{p-1}^2)^\perp}\,\Pi^{(p-1)}\,
\delta^{(d)}(x,y)\nonumber\\\left\{\gamma^k\,,\,\overline{\gamma}^l\right\}
&=&-\frac{2\pi i}k\,(-1)^{p(d-p)}\,G^{kl}
\label{poisson}\eea
where $\Pi^{(p)}:\Omega^p(\MD)\to\Omega^p(\MD)$ is the symmetric,
transverse orthogonal projection onto the subalgebra of co-closed $p$-forms of
$\Omega^p(\MD)$, and $(\nabla_p^2)^\perp$ denotes the Laplacian operator
$\nabla_p^2$ with its zero modes arising from gauge invariance removed.

In the
quantum field theory, Poisson brackets are mapped onto quantum commutators
according to the correspondence principle. In the functional Schr\"odinger
picture, we may treat the fields $\theta$, $K$ and $\gamma^k$ as
``coordinates'' on the (infinite dimensional) configuration space \cite{djt}. Then the
canonical commutation relations corresponding to \eqn{poisson} are represented by
writing the canonical momentum differential forms as the derivative operators
\bea
P_\theta(x)&=&\frac{2\pi i}k\,\frac1{(\nabla_{d-p-1}^2)^\perp}\,
\Pi^{(d-p-1)}\,\frac\delta{\delta\theta(x)}\nonumber\\P_K(x)&=&\frac{2\pi
i}k\,\frac1{(\nabla_{p-1}^2)^\perp}\,\Pi^{(p-1)}\,\frac\delta{\delta
K(x)}\nonumber\\\overline{\gamma}^l&=&\frac{2\pi}k\,(-1)^{p(d-p)}\,G^{lk}
\,\frac\partial{\partial\gamma^k}
\label{canmomreps}\eea
As we shall see, the projection operators in \eqn{canmomreps} have the effect 
of ensuring the invariance of the physical state wavefunctions under the
time-independent secondary gauge symmetries
\beq
\theta\to\theta+d\chi''~~~~~~,~~~~~~K\to K+d\xi''
\label{secondarygts}\eeq
These symmetries are a consequence of the feature that the topological gauge
theory has first-stage, off-shell reducible gauge symmetries \cite{bt,tqft}, and they can
be regarded, through the minimal couplings of the gauge fields to the
deRham currents, as being dual to the current symmetries \eqn{currentsym}.

Finally, we can express the Hamiltonian \eqn{ham} as well in terms of the
various Hodge decompositions of the previous subsection. Using the fact that the
singular parts of the deRham currents make no contributions, after some algebra
and integrations by parts we arrive at the classical Hamiltonian
\bea
H&=&-\int\limits_{\MD}\left((-1)^d\,Q_{d-p}\,\theta\wedge*\nabla_{d-p-1}^2
\omega_{d-p}-(-1)^{(d-1)(p-1)}Q_p\,K\wedge*\nabla_{p-1}^2\omega_p
\right.\nn\\& &\left.+\,(-1)^d\,Q_{d-p}\,*\rho_{d-p}\wedge\nabla_{p-1}^2P_K
-(-1)^{(d-1)(p-1)}\,Q_p\,*\rho_p\wedge\nabla_{d-p-1}^2P_\theta\right)
\nonumber\\& &+\,(-1)^{(d-1)p}\,i\sum_{m=1}^{N_p}\left(Q_p\,\Sigma_m^{(p)}
-Q_{d-p}\sum_{n,l=1}^{N_p}\overline{\tau}_{mn}I^{(p)nl}\,\Sigma_l^{(d-p)}
\right)\sum_{r,k=1}^{N_p}I^{(p)mr}G_{rk}\,\gamma^k\nonumber\\& &
-\,(-1)^{(d-1)p}\,i\sum_{m=1}^{N_p}\left(Q_p\,\Sigma_m^{(p)}-Q_{d-p}
\sum_{n,l=1}^{N_p}\tau_{mn}I^{(p)nl}\,\Sigma_l^{(d-p)}\right)
\sum_{r,k=1}^{N_p}I^{(p)mr}G_{rk}\,\overline{\gamma}^k
\nn\\& &
\label{classham}\eea
Using the continuity equations \eqn{conteq} and the Schr\"odinger
representations \eqn{canmomreps}, we arrive at the corresponding quantum
Hamiltonian operator:
\bea
H&=&-\int\limits_{\MD}\left(-(-1)^d\,Q_{d-p}\,\theta\wedge*
\,\frac\partial{\partial
t}\,\bar\delta^0_{\Sigma_{d-p}}-(-1)^{(d-1)(p-1)}\,Q_p\,K\wedge*
\,\frac\partial{\partial
t}\,\bar\delta^0_{\Sigma_p}\right.\nonumber\\& &\left.+\,\frac{2\pi
i}k\,Q_{d-p}*\,\rho_{d-p}\wedge\Pi^{(p-1)}\,\frac\delta{\delta K}+\frac{2\pi
i}k\,Q_p\,*\rho_p\wedge\Pi
^{(d-p-1)}\,\frac\delta{\delta\theta}\right)\nonumber\\&
&+\,(-1)^{(d-1)p}\,i\sum_{m=1}
^{N_p}\left(Q_p\,\Sigma_m^{(p)}-Q_{d-p}\sum_{n,l=1}^{N_p}\overline{\tau}_{mn}
I^{(p)nl}\,\Sigma_l^{(d-p)}\right)\sum_{r,k=1}^{N_p}
I^{(p)mr}G_{rk}\,\gamma^k\nonumber\\& &-\,\frac{2\pi i}k\sum_{m=1}
^{N_p}\left(Q_p\,\Sigma_m^{(p)}-Q_{d-p}\sum_{n,l=1}^{N_p}\tau_{mn}
I^{(p)nl}\,\Sigma_l^{(d-p)}\right)\sum_{k=1}^{N_p}
I^{(p)mk}\,\frac\partial{\partial\gamma^k}
\label{quantham}\eea

\newsubsection{Global Phase Space}

We close this section by comparing the present canonical approach to the path
integral formalism used in \cite{newbf} for the quantization of abelian $BF$
theory defined over non-trivial line bundles. In the former approach, the zero
mode vector space \eqn{calPdef} comes from the single-valued parts of the gauge
fields and would yield the same overall volume factor in the path integral
measure as that for the quantum field theory defined on a trivial vector
bundle. The sum over all line bundles localizes onto the topological class
determined by the external charges $Q_p$ and $Q_{d-p}$ through the flux
relationships \eqn{FQrel} and \eqn{HQrel}.
The non-triviality of the bundle is then encoded in the strong gauge
constraints \eqn{topquant} which will be imposed on the wavefunctions in the
next section. The topological sum may then be taken over all quantized charges
obeying \eqn{topquant}.
In the latter approach, the sum over non-trivial line bundles is
explicitly carried out in the source-free
path integral and is shown to modify the (ungraded) space
of harmonic zero modes to
\beq
\check{\cal P}=\bigoplus_{n=p,d-p}\left(H_{\rm C}^n(\MD,\zed)\,/\,H_{\rm C}^{n-1}
(\MD,\IR/\zed)\right)\oplus\left(H_{\rm C}^{d-n}(\MD,\IR/\zed)\otimes
H_{\rm D}^{d-n}(\MD)\right)
\label{Pcech}\eeq
Here the \v{C}ech cohomology group $H_{\rm C}^{d-p}(\MD,\IR/\zed)$ classifies
the higher-rank bundles of degree $d-p$ which admit flat $d-p$-form connections
(equivalently constant transition functions), while the quotient cohomology
in \eqn{Pcech} classifies the higher-rank bundles of degree $d-p$ modulo those
with constant transition functions. Thus, although the functional integration
reproduces the usual Ray-Singer invariant, the overall volume factor is now
modified to take into account of the non-trivial \v{C}ech cohomology. In the
following we shall see how this cohomology is represented explicitly by the
wavefunctions of the canonical quantum field theory.

\newsection{Construction of the Physical States}

In this section we shall use the canonical formalism for the quantum field
theory developed in the previous section to explicitly solve for the
physical state wavefunctions of $BF$ theory. From the form of the Hamiltonian
operator \eqn{quantham} we see that they may be separated into 
two pieces $\Psi_L$ and $\Psi_T$ representing the local and global topological
parts in the corresponding decomposition of the $BF$ field theory:
\beq
\Psi_{\rm phys}[\theta,K,\gamma;t]=\Psi_L[\theta,K;t]\,\Psi_T(\gamma;t)
\label{Psiphyscomp}\eeq
We shall see that each component in \eqn{Psiphyscomp} plays an important role in
the topological group representations that we will find in the next section.

\newsubsection{Local Gauge Symmetries}

The local components of the full wavefunction must satisfy the weak equalities
\eqn{Aweakeq} and \eqn{Bweakeq} which are imposed
as physical state conditions in the quantum field theory and which truncate the
full Hilbert space onto the physical, gauge invariant
subspace. Using the functional Schr\"odinger
representations in \eqn{canmomreps}, we thereby arrive at the 
quantum equations which
express the local gauge constraints of the theory:
\bea
\left[(-1)^{(d-1)(p-1)}\,i\,\Pi^{(p-1)}\,\frac\delta{\delta K}+Q_p\,\bar\delta
_{\Sigma_p}^0\right]\Psi_L[\theta,K;t]&=&0\nn\\\left[-(-1)^d\,i\,\Pi^{(d-p-1)}\,
\frac\delta{\delta\theta}+Q_{d-p}\,\bar\delta^0_{\Sigma_{d-p}}\right]
\Psi_L[\theta,K;t]&=&0
\label{wavelocconstr}\eea
They are solved by wavefunctionals of the form
\beq
\Psi_L[\theta,K;t]=\exp\left[i\int\limits_{\MD}\left((-1)^{(d-1)(p-1)}\,Q_p
\,K\wedge*\bar\delta_{\Sigma_p}^0-(-1)^d\,Q_{d-p}\,\theta\wedge*\bar
\delta^0_{\Sigma_{d-p}}\right)\right]\,\tilde\Psi_L(t)
\label{locgaugesoln}\eeq
which yield a projective representation of the local gauge symmetries in 
terms of a non-trivial, local $U(1)\times U(1)$ one-cocycle:
\bea
& &\Psi_{\rm phys}[\theta+\chi',K+\xi',
\gamma;t]\nn\\& &=\exp\left[i\int\limits_{\MD}
\left((-1)^{(d-1)(p-1)}\,Q_p\,\xi'\wedge*\bar\delta_{\Sigma_p}^0-(-1)^d\,Q_{d-p}
\,\chi'\wedge*\bar\delta_{\Sigma_{d-p}}^0\right)\right]\,\Psi_{\rm
phys}[\theta,K,\gamma;t]
\nn\\& &
\label{locprojrep}\eea
Note that
the wavefunction \eqn{locgaugesoln} can be written in a more explicit form using
the decompositions \eqn{Dsingdecomp} of the deRham currents.

\newsubsection{Schr\"odinger Wave Equation}

To determine the remaining part $\tilde\Psi_L(t)$ of the wavefunction
\eqn{locgaugesoln}, we solve the corresponding Schr\"odinger wave equation
\beq
i\frac\partial{\partial t}\Psi_{\rm phys}[\theta,K,\gamma;t]=H\Psi_{\rm phys}
[\theta,K,\gamma;t]
\label{waveeq}\eeq
for the local degrees of freedom of the gauge fields. From \eqn{quantham} we 
readily arrive at
\beq
\tilde\Psi_L(t)=\exp-\frac{2\pi i}k\,Q_p\,Q_{d-p}\,(-1)^{p(d-1)}\int
\limits_0^tdt'~\int\limits_{\MD}\left(-*\rho_{d-p}\wedge\bar\delta^0_{\Sigma_p}
+*\rho_p\wedge\bar\delta^0_{\Sigma_{d-p}}\right)
\label{tildePsisoln}\eeq
Substituting in the eigenfunction expansions \eqn{Deigen} gives
\bea
\tilde\Psi_L(t)&=&\exp\Biggl\{-\frac{2\pi i}k\,Q_p\,Q_{d-p}\,(-1)^{p(d-1)}
\Biggr.\nn\\& &\times
\int\limits_0^tdt'~\left[-\sum_{\lambda_{d-p}\neq0}\left(\frac d{dt'}\int
\limits_{\Sigma_p(t')}\psi_{\lambda_{d-p}}^{(d-p)}\right)\left(
\int\limits_{\Sigma_p(t')}*\psi_{\lambda_{d-p}}^{(d-p)}\right)\right.
\nn\\
& &\left.
\left.+\sum_{\lambda_{d-p-1}\neq0}\frac1{\lambda_{d-p-1}^2}\left(\frac d{dt'}
\int\limits_{\Sigma_p(t')}*d\psi_{\lambda_{d-p-1}}^{(d-p-1)}\right)
\left(\oint\limits_{\partial\Sigma_{d-p}(t')}\psi_{\lambda_{d-p-1}}^{(d-p-1)
}\right)\right]\right\}\nn\\& &
\nn\\&=&\exp\Biggl\{-\frac{2\pi i}k\,Q_p\,Q_{d-p}
\,(-1)^{p(d-1)}\Biggr.\nn\\& &\times\int\limits_0^tdt'~\left[\frac d{dt'}
\sum_{\lambda_{d-p-1}\neq0}\frac1{\lambda_{d-p-1}^2}\left(\int\limits_{
\Sigma_p(t')}*d\psi_{\lambda_{d-p-1}}^{(d-p-1)}\right)\left(\oint\limits_{
\partial\Sigma_{d-p}(t')}\psi_{\lambda_{d-p-1}}^{(d-p-1)}\right)\right.
\nn\\& &-\sum_{l,m=1}^{N_p}\left(\int\limits_{\Sigma_p(t')}\beta_l^{(p)}\right)
I^{(p)lm}\left(\frac d{dt'}\int\limits_{\Sigma_{d-p}(t')}\alpha_m^{(p)}
\right)\nn\\& &-\sum_{\lambda_{d-p-1}\neq0}\frac1{\lambda_{d-p-1}^2}
\left(\int\limits_{\Sigma_p(t')}*d\psi_{\lambda_{d-p-1}}^{(d-p-1)}
\right)\left(\oint\limits_{\partial\Sigma_{d-p}(t')}\imath_{\dot X_{d-p}}d
\psi_{\lambda_{d-p-1}}^{(d-p-1)}\right)\nn\\& &+\sum_{\lambda_{d-p}\neq0}
\left(\int\limits_{\Sigma_p(t')}*\psi_{\lambda_{d-p}}^{(d-p)}\right)
\left(\oint\limits_{\partial\Sigma_{d-p}(t')}\imath_{\dot X_{d-p}}
\psi_{\lambda_{d-p}}^{(d-p)}\right)\nn\\& &
\left.\left.+\sum_{l,m=1}^{N_p}\left(\int
\limits_{\Sigma_p(t')}\beta_l^{(p)}\right)I^{(p)lm}\left(\oint
\limits_{\partial\Sigma_{d-p}(t')}\imath_{\dot X_{d-p}}\alpha_m^{(p)}
\right)\right]\right\}
\label{tildePsieigen}\eea
where $\imath_{\dot X_{d-p}}:\Omega^q(\MD)\to\Omega^{q-1}(\MD)$ is the nilpotent
interior multiplication with respect to the vector field $\frac d{dt}X_{d-p}^\mu
(t,\sigma^2,\dots,\sigma^{d-p})$. From \eqn{tildePsieigen} we arrive finally at
\bea
\tilde\Psi_L(t)&=&\exp\Biggl\{-\frac{2\pi i}k\,Q_p\,Q_{d-p}\,(-1)^{p(d-1)}
\Biggr.\nn\\& &\left.
\times\int\limits_0^tdt'~\left[\frac1{\Omega_{d-1}}
\,\frac{d\Phi_p(t')}{dt'}+\sum_{l,m=1}^{N_p}\Sigma_m^{(d-p)}(t')
I^{(p)lm}\int\limits_0^{t'}dt''~\Sigma_l^{(p)}(t'')\right]\right\}
\label{tildePsiPhi}\eea
where we have introduced the function
\bea
\Phi_p(t)&=&\Omega_{d-1}\int\limits_0^tdt'~\left[
\oint\limits_{\partial\Sigma_{d-p}(t')}\int\limits_{\Sigma_p(t')}
\left(\imath_{\dot X_{d-p}}\otimes1\right)\,\delta^{(d-p,p)}
\left(X_{d-p}(t',\sigma^\alpha)\,,\,X_p(t',\sigma'^\beta
)\right)\right]\nn\\& &+\,\Omega_{d-1}
\sum_{\lambda_{d-p-1}\neq0}\frac1{\lambda_{d-p-1}^2}\left(
\oint\limits_{\partial\Sigma_{d-p}(t)}\psi_{\lambda_{d-p-1}}^{(d-p-1)}
\right)\left(\int\limits_{\Sigma_p(t)}*d\psi_{\lambda_{d-p-1}}^{(d-p-1)}
\right)
\label{Phidef}\eea
and
\beq
\Omega_{d-1}={\rm vol}({\bf S}^{d-1})=\frac{2\pi^{d/2}}{\Gamma(\mbox{$\frac d2$})}
\label{solidd}\eeq
is the $d-1$-dimensional solid angle.
Note that the above derivation can be carried out in the same way to obtain a final
expression which is explicitly symmetric in $\Sigma_{d-p}(t)$ and $\Sigma_p(t)$,
and which thereby exhibits the manifest (Hodge) duality symmetry between the two
hypersurfaces. However, in order to keep the formulas from getting overly lengthy,
we do not exhibit this symmetrization explicitly.

Let us examine the dependence of the function \eqn{Phidef} on the topological
classes of the projected hypersurfaces. For this, we fix the $p-1$-brane
embedding function $X_p(t,\sigma^2,\dots,\sigma^p)$ and choose another
hypersurface $\Sigma_p'(t)$ in the same topological class as $\Sigma_p(t)$,
i.e. $\Sigma_p(t)-\Sigma_p'(t)=\partial{\cal S}_p(t)$ for some $p+1$-volume
${\cal S}_p(t)$. It is then straightforward to compute the change in 
the second term in \eqn{Phidef}:
\bea
\delta\Phi_p(t)&=&\Omega_{d-1}\sum_{\lambda_{p-1}\neq0}
\frac1{\lambda_{d-p-1}^2}\left(\oint\limits_{\partial\Sigma_{d-p}(t)}
\psi_{\lambda_{d-p-1}}^{(d-p-1)}\right)\left(\oint\limits_{\partial
{\cal S}_p(t)}*d\psi_{\lambda_{d-p-1}}^{(d-p-1)}\right)\nn\\&=&
\Omega_{d-1}\left[\oint\limits_{\partial\Sigma_{d-p}(t)}\int\limits
_{{\cal S}_p(t)}\delta^{(d-p-1,p+1)}\left(X_{d-p}(t,\sigma^2,\dots,\sigma
^{d-p})\,,\,s_p(t)\right)\right.\nn\\& &\left.-\sum_{l,m=1}^{N_{p+1}}\left(
\oint\limits_{\partial\Sigma_{d-p}(t)}\alpha_m^{(p+1)}\right)I^{(p+1)lm}
\left(\int\limits_{{\cal S}_p(t)}\beta_l^{(p+1)}\right)\right]
\label{deltaPhi}\eea
where we have used Stokes' theorem, and $s_p(t)$ is a local coordinate 
system on ${\cal S}_p(t)$. This shows that if we continuously deform 
$\Sigma_p(t)$ in $\MD$, then, modulo the harmonic forms in \eqn{deltaPhi},
the second term in \eqn{Phidef} does not change unless the deformation 
crosses the $d-p-1$-brane at $X_{d-p}(t,\sigma^2,\dots,\sigma^{d-p})$.
In that case the change is then $\Omega_{d-1}$, which is cancelled
by the delta-function term in \eqn{Phidef}. As for the harmonic part of
\eqn{deltaPhi}, we symmetrize the expression \eqn{tildePsiPhi} in $\Sigma_p(t)$
and $\Sigma_{d-p}(t)$ to obtain a manifestly Hodge duality symmetric function.
Then the harmonic term in \eqn{deltaPhi} becomes
\bea
& &-\Omega_{d-1}\left[\sum_{l,m=1}^{N_{p+1}}\left(
\oint\limits_{\partial\Sigma_{d-p}(t)}\alpha_m^{(p+1)}I^{(p+1)lm}\right)
\left(\int\limits_{{\cal S}_p(t)}\beta_l^{(p+1)}\right)
\right.\nn\\& &~~~~~~~~~~~~\left.
+\sum_{l,m=1}^{N_{p-1}}\left(\oint\limits_{\partial\Sigma_p(t)
}\beta_m^{(p-1)}\right)I^{(p-1)ml}\left(\int\limits_{{\cal S}_{d-p}(t)}
\alpha_m^{(d-p)}\right)\right]
\label{symharmchange}\eea
Using the bilinear identity \eqn{pairing1}, we see that the change
\eqn{symharmchange} will contribute a phase factor to \eqn{tildePsiPhi} 
which is simply unity due to the topological phase constraint \eqn{topquant}
(applied to each time slice of $\MDD$). The condition \eqn{topquant}
thereby represents a fundamental global constraint that must be satisfied by
the external charges $Q_p,Q_{d-p}$ for a consistent (topologically
invariant) solution of the $BF$ quantum field theory. We shall see later
on that this imposes a corresponding global constraint that must be met by all
consistent well-defined representations of the motion group on topologically
non-trivial manifolds $\MD$. 

The function $\Phi_p(t)$ therefore depends only on the topological classes
of the trajectories $\Sigma_p(t)$ and $\Sigma_{d-p}(t)$ in $\MD$, and not
on their particular representatives, provided that they do not intersect.
The above argument also shows that if $\Sigma_{d-p}(t)$ is kept constant in
time $t$ while $\Sigma_p(t)$ sweeps out a closed hypersurface in a given
time span, then the only contribution to $\Phi_p(t)$ is from the second term
in \eqn{Phidef} which gives $\Omega_{d-1}$. Furthermore, if $\Sigma_p(t)$
is fixed and $\Sigma_{d-p}(t)$ sweeps out a closed hypervolume in a given 
time span, then the second term in \eqn{Phidef} is invariant while the first 
term counts exactly the number of times the hypersurface $\Sigma_{d-p}(t)$ 
links $\Sigma_p(t)$, giving a contribution of $\Omega_{d-1}$ each
time. Thus, $\Phi_p(t)$ gives the $d$-dimensional relative solid angle
between $\Sigma_p(t)$ and $\Sigma_{d-p}(t)$ in adiabatic linking processes
in $\MD$. It is the generalized, adiabatic linking function
that will yield the appropriate holonomy phase changes in the wavefunctions
for the motion group representations that we will obtain in the following.
Furthermore, for infinitesimal paths it is readily seen that \eqn{Phidef} reduces
to the usual solid angle function on $\MD=\IR^d$. We will describe the
transformation properties of the wavefunctions under homologically 
non-trivial motions of the hypersurfaces $\Sigma_p(t)$ and $\Sigma_{d-p}(t)$
later on.

We now come to the remaining, topological part of the Schr\"odinger equation
\eqn{waveeq} for the global harmonic degrees of freedom of the gauge
fields. From \eqn{quantham} it follows that this equation may be solved
in the form
\bea
\Psi_T(\gamma;t)&=&\prod_{m,n,r=1}^{N_p}\exp\left[(-1)^{p(d-1)}\int
\limits_0^tdt'~\left(Q_p\,\Sigma_m^{(p)}(t')-Q_{d-p}\sum_{s,l=1}^{N_p}
\overline{\tau}_{ms}I^{(p)sl}\,\Sigma_l^{(d-p)}(t')\right)
\right.\nn\\& &\times\,I^{(p)mr}
G_{rn}\,\gamma^n\nn\\& &-\frac{2\pi}k\,(-1)^{p(d-1)}\int\limits
_0^tdt'~\left(Q_p\,\Sigma_m^{(p)}(t')-Q_{d-p}\sum_{u,l=1}^{N_p}
\tau_{mu}I^{(p)nl}\,\Sigma_l^{(d-p)}(t')\right)\nn\\& &\left.\times
\sum_{q=1}^{N_p}
I^{(p)mq}\,G_{qn}\,I^{(p)rn}\int\limits_0^{t'}dt''~
\left(Q_p\,\Sigma_r^{(p)}(t'')-Q_{d-p}\sum_{s,v=1}^{N_p}
\overline{\tau}_{rv}I^{(p)vs}\,\Sigma_s^{(d-p)}(t'')\right)\right]
\nn\\& &\times\,\Psi_0(\gamma;t)
\label{PsiTform}\eea
where the function $\Psi_0(\gamma;t)$ is a solution of the partial differential
equation
\bea
\frac{\partial\Psi_0(\gamma;t)}{\partial t}&=&-\frac{2\pi}k\,(-1)^{p(d-1)}
\sum_{m=1}^{N_p}\left(Q_q\,\Sigma_m^{(p)}(t)-Q_{d-p}\sum_{n,l=1}^{N_p}
\tau_{mn}I^{(p)nl}\,\Sigma_l^{(d-p)}(t)\right)\nn\\& &
\times\sum_{r=1}^{N_p}I^{(p)mr}\,
\frac{\partial\Psi_0(\gamma;t)}{\partial\gamma^r}
\label{Psi0pde}\eea
which is solved by any function of the form
\bea
& &\Psi_0(\gamma^l;t)\nn\\& &
=\Psi_0\left(\mbox{$\gamma^l-\frac{2\pi}k\,(-1)^{p(d-1)}
\sum_mI^{(p)ml}\int_0^tdt'~\left[Q_p\,\Sigma^{(p)}_m(t')-Q_{d-p}
\sum_{n,r}\tau_{mn}I^{(p)nr}\,\Sigma_r^{(d-p)}(t')\right]$}\right)
\nn\\& &
\label{Psi0form}\eea
The function $\Psi_0$ may be fixed by requiring that the wavefunctions 
respect the large gauge transformations of the fields which are not
connected to the identity in the topological phase space \eqn{calPdef}.
This will be done in the next subsection.

\newsubsection{Global Gauge Symmetries}

For a consistent quantum theory, we must demand that, when there are no
sources present ($Q_p=Q_{d-p}=0$), the wavefunctions $\Psi_0$ coincide with
the cohomological states that represent the invariance of the quantum
field theory under large gauge transformations. In this case, the local
gauge constraints \eqn{wavelocconstr} imply that the full physical state 
wavefunctions depend only on the global harmonic degrees of freedom $\gamma^l$.
Furthermore, the Hamiltonian then vanishes (since the pure source-free
$BF$ field theory is topological) so that the states are also 
time-independent. This means that in the absence of any sources the
wavefunctions carry information only about the topology of the
manifold $\MD$.

To construct these states, we consider the classical translation operators
which generate the appropriate shifts \eqn{largegaugegamma} of the 
holomorphic gauge degrees of freedom:
\bea
C(n_p,n_{d-p})&=&\prod_{l=1}^{N_p}\exp\left[2\pi\left(
n_p^l+\sum_{m,r=1}^{N_p}
I^{(p)ml}\tau_{mr}n_{d-p}^r\right)\frac\partial{\partial\gamma^l}
\right.\nn\\& &\left.+\,2\pi\left(n_p^l+\sum_{m,r=1}^{N_p}I^{(p)ml}\overline{
\tau}_{mr}n_{d-p}^r\right)\frac\partial{\partial\overline{\gamma}^l}
\right]
\label{classgens}\eea
Using the Schr\"odinger representation \eqn{canmomreps} we can 
then write down the
corresponding quantum operators which implement the global gauge symmetries:
\bea
U(n_p,n_{d-p})&=&\prod_{l=1}^{N_p}\exp\left[2\pi\left(n_p^l+
\sum_{m,r=1}^{N_p}I^{(p)ml}\tau_{mr}n_{d-p}^r\right)\frac
\partial{\partial\gamma^l}\right.\nn\\& &\left.-\,(-1)^{p(d-p)}\,k
\left(n_p^l+\sum_{m,r=1}^{N_p}I^{(p)ml}\overline{\tau}_{mr}
n_{d-p}^r\right)\sum_{q=1}^{N_p}G_{lq}\,\gamma^q\right]
\label{quantgens}\eea
For the remainder of this paper we will assume that the coefficient
$k$ of the pure $BF$ action is of the form
\beq
k={\cal I}^{(p)}\,\frac{k_1}{k_2}
\label{coeffquant}\eeq
where ${\cal I}^{(p)}>0$ is the integer-valued determinant of the 
intersection matrix $I^{(p)lm}$ and $k_1,k_2$ are positive integers with
${\rm gcd}({\cal I}^{(p)}k_1,k_2)=1$.

In contrast to the classical operators \eqn{classgens}, the operators 
$U(n_p,n_{d-p})$ do not commute with each other. Using the 
Baker-Campbell-Hausdorff formula it is straightforward to compute that
these operators generate the global $U(1)\times U(1)$ two-cocycle algebra:
\beq
U(n_p,n_{d-p})U(m_p,m_{d-p})=\left[\prod_{l,r=1}^{N_p}\e^{2\pi
ik(-1)^{p(d-p)}I_{lr}^{(p)}\left(m_p^ln_{d-p}^r-n_p^lm_{d-p}^r\right)}\right]
\,U(m_p,m_{d-p})U(n_p,n_{d-p})
\label{clockalg}\eeq
A similar calculation shows that their action on the
wavefunctions is given by
\bea
U(n_p,n_{d-p})\Psi_0(\gamma^l)&=&\prod_{l,q=1}^{N_p}\exp(-1)^{p(d-p)}
\left\{-k\left(n_p^l+\sum_{m,r=1}^{N_p}I^{(p)ml}\overline{\tau}_{mr}
n_{d-p}^r\right)G_{lq}\,\gamma^q\right.
\nn\\& &\left.-\,\pi k\left(n_p^l+\sum_{m,r=1}
^{N_p}I^{(p)ml}\overline{\tau}_{mr}n_{d-p}^r\right)G_{lq}\left(n_p^q+
\sum_{s,u=1}^{N_p}I^{(p)sq}\tau_{su}n^u_{d-p}\right)\right\}\nn\\& &
\times\,\Psi_0\left(\mbox{$\gamma^l+2\pi(n_p^l+\sum_{m,r}I^{(p)ml}
\tau_{mr}n_{d-p}^r)$}\right)
\label{UPsiaction}\eea
On the other hand, the cocycle algebra \eqn{clockalg} implies that the
operators $U(k_2n_p,k_2n_{d-p})$ commute with all of the other gauge 
transformation generators, so that they lie in the center of the global
$U(1)\times U(1)$ gauge group and their action on the Hilbert space
is represented simply as multiplication by some phases $\e^{i\phi
(n_p,n_{d-p})}$. This then implies the transformation law:
\bea
& &\Psi_0\left(\mbox{$\gamma^l+2\pi k_2(n_p^l+\sum_{m,r}I^{(p)ml}
\tau_{mr}n_{d-p}^r)$}\right)\nn\\& &~~~=\exp\left[i\phi(n_p,n_{d-p})
+\sum_{l=1}^{N_p}(-1)^{p(d-p)}\left\{k_1\left(n_p^l+\sum_{m,r=1}^{N_p}
I^{(p)ml}\overline{\tau}_{mr}n_{d-p}^r\right)
\sum_{q=1}^{N_p}G_{lq}\,\gamma^q\right.\right.
\nn\\& &\left.\left.~~~~~~+\,\pi k_1k_2
\left(n_p^l+\sum_{m,r=1}^{N_p}I^{(p)ml}\overline{\tau}_{mr}
n_{d-p}^r\right)\sum_{q=1}^{N_p}
G_{lq}\left(n_p^q+\sum_{s,u=1}^{N_p}I^{(p)sq}\tau_{su}n_{d-p}^u\right)
\right\}\right]\,\Psi_0(\gamma^l)\nn\\& &
\label{transflaw}\eea
These algebraic constraints are uniquely solved by the $({\cal I}^{(p)}
k_1k_2)^{N_p}$ independent holomorphic wavefunctions
\bea
\Psi_0^{(q)}{a\choose b}(\gamma_l)&=&\left(\prod_{l,r=1}^{N_p}\e^{\frac k{4\pi}
\gamma^lG_{lr}\gamma^r}\right)\nn\\& &\times\,\Theta
{\frac{a+q}{{\cal I}^{(p)}k_1k_2}\choose b}\left(\left.\frac{{\cal I}^{(p)}
k_1}{2\pi}\,(-1)^{p(d-p)}\sum_{m=1}^{N_p}I^{(p)}_{ml}\,\gamma^m
\right|-k_1k_2{\cal I}^{(p)}\tau\right)
\label{uniquelargesoln}\eea
where $q^l=1,2,\dots,{\cal I}^{(p)}k_1k_2$ ($l=1,\dots,N_p$), and we have
introduced the standard (multi-dimensional) Jacobi theta-functions:
\beq
\Theta{a\choose b}(z|-\tau)=\sum_{n\in\Gamma}\prod_{l=1}^{{\rm rank}(\Gamma)}
\exp\left[-i\pi\left(n^l+a^l\right)\sum_{k=1}^{{\rm rank}(\Gamma)}
\tau_{lk}\left(n^k+a^k\right)+2\pi(n^l+a^l)(z_l+b_l)\right]
\label{Jacobidef}\eeq
where $a^l,b_l\in[0,1]$ and $\Gamma$ is some lattice. The functions
\eqn{Jacobidef} are well-defined and holomorphic in $z\in\complex^{{\rm
rank}(\Gamma)}$ when the lattice quadratic form $-\tau$ is an element of
the Siegal upper half-plane.

In the present case, the lattice $\Gamma$ is taken to be the torsion-free
part of the integer \v{C}ech cohomology group $H_{\rm C}^p(\MD,\zed)$, and
the corresponding dual lattice $\Gamma^*$ that of $H_{\rm C}^{d-p}
(\MD,\zed)$, of rank $N_p$. Then the standard double semi-periodicity
properties of the Jacobi theta-functions \cite{bss,mumford}
provides the unique solution
\eqn{uniquelargesoln} of the quasi-periodicity constraints
\eqn{transflaw}. The wavefunctions \eqn{uniquelargesoln} are orthogonal in 
the canonical coherent state measure on the reduced topological phase
space ${\cal P}/\Gamma\oplus\Gamma^*$ which leads to the inner product:
\bea
\left(\Psi_0^{(q)}\left|\Psi_0^{(q')}\right.\right)&=&\int\limits
_{{\cal P}/\Gamma\oplus\Gamma^*}\prod_{m=1}^{N_p}d\gamma^m~d\overline{
\gamma}^m~\prod_{k,l=1}^{N_p}\e^{-\frac k{2\pi}\gamma^kG_{kl}\gamma^l}\,
(\det G)^{-1}\,\Psi_0^{(q)}(\overline{\gamma})^*\,\Psi_0^{(q')}(\gamma)
\nn\\&=&(\det G)^{-1/2}\,\delta^{qq'}
\label{cohstatemeas}\eea
where we have implicitly divided out by the volume of the global gauge
group used to define the complex $N_p$-torus ${\cal P}/\Gamma\oplus
\Gamma^*$ (as a consequence of the large gauge invariances). The states
\eqn{uniquelargesoln} thereby provide a complete, orthonormal basis of
the full physical Hilbert space, and they are well-defined functions on
${\cal P}/\Gamma\oplus\Gamma^*$. Furthermore, under a large gauge 
transformation these wavefunctions transform as
\beq
U(n_p,n_{d-p})\Psi_0^{(q)}{a\choose b}(\gamma)=\sum_{q'}U(n_p,n_{d-p})_{qq'}
\,\Psi_0^{(q')}{a\choose b}(\gamma)
\label{Psi0transfU}\eeq
where the unitary matrices
\bea
U(n_p,n_{d-p})_{qq'}&=&\prod_{l,m=1}^{N_p}\exp\Biggl\{\frac{2\pi
i}{k_2}\,(-1)^{p(d-p)}I_{ml}^{(p)}\Biggr.\nn\\& &\left.
\times\left(a^ln_p^m+\sum_{r=1}^{N_p}I^{(p)lr}b_rn_{d-p}^l+q^ln_p^m-
\frac{{\cal I}^{(p)}k_1}2\,n_p^mn_{d-p}^l\right)\right\}
\nn\\& &\times\,\delta_{q^r-k_1{\cal I}^{(p)}n_{d-p}^r\,,\,q'^r}
\label{unitarytransf}\eea
generate a $(k_2)^{N_p}$-dimensional projective representation of the
group $\Gamma\oplus\Gamma^*$ of large gauge transformations. Here the
projective phases are non-trivial global $U(1)\times U(1)$ one-cocycles
which are cyclic with period $k_2$. The topological part of the full
wavefunction thereby carries a non-trivial multi-dimensional projective
representation of the discrete gauge group representing the windings
of the $BF$ gauge fields around the appropriate non-trivial homology cycles
of $\MD$. This symmetry partitions the Hilbert space into superselection
sectors labelled by the integer (\v{C}ech) cohomology classes of the
spatial manifold.

The wavefunctions \eqn{uniquelargesoln} possess some noteworthy 
modular transformation properties. The automorphism group of the
reduced topological phase space ${\cal P}/\Gamma\oplus\Gamma^*$ with
its complex structure $\tau$ and associated metric $G$ is
$Sp(2N_p,\zed)$. It acts on the geometrical parameters as
\bea
\gamma'&=&(-C\tau+D)^{-1\,\top}\,\gamma\nn\\\tau'&=&-(-A\tau+B)(-C\tau+D)^{-1}
\nn\\G'&=&(-C\tau+D)^{-1\,\top}\,G\,(-C\tau+D)^{-1}
\label{autoaction}\eea
where ${\scriptstyle\addtolength{\arraycolsep}{-.5\arraycolsep}
\renewcommand{\arraystretch}{0.5}\left(\begin{array}{cc}
\scriptstyle A & \scriptstyle B \\ \scriptstyle C &
\scriptstyle D \end{array}\scriptstyle\right)}\in Sp(2N_p,\zed)$. The
corresponding transformation of the Jacobi theta-functions \eqn{Jacobidef}
is given by \cite{mumford}
\beq
\Theta{a'\choose b'}(\gamma'|-\tau')=\e^{-i\pi\phi}\,\sqrt{\det(-C\tau+D)}
~\e^{i\pi\gamma^\top(-C\tau+D)^{-1}\gamma}\,\Theta{a\choose b}(\gamma|-\tau)
\label{Jacobitransf}\eeq
where $\phi$ is an irrelevant phase and
\beq
a'=Da-Cb-\mbox{$\frac12$}\,(CD^\top)_{\rm diag}~~~~~~,~~~~~~b'=-Ba+Ab-
\mbox{$\frac12$}\,(AB^\top)_{\rm diag}
\label{parchange}\eeq
It follows that a modular invariant set of wavefunctions exists only when
the quantity ${\cal I}^{(p)}k_1k_2$ is an even integer, in which case
we set $a^l=b_l=0$ (and also $\phi=0$). Otherwise, we may take
$a^l,b_l\in\{0,\frac12\}$, which corresponds to a choice of spin structure
on the complex $N_p$-torus which is the reduced topological phase space. 
The totality of wavefunctions labelled by the $a^l,b_l$ then increases by
$4^{N_p}$, and they now transform non-trivially under modular transformations
representing their transformation properties under a change in choice of
spin structure. These latter choices of $a^l,b_l$ are the ${\it only}$ ones
for which the reflection symmetry $\gamma\to-\gamma$ closes on the set of
wavefunctions \eqn{uniquelargesoln} \cite{bss}. In this way, the physical states turn
out to be effectively independent of the phase space complex structure, as 
required by the topological nature of the quantum field theory.

\newsection{Representations of Motion Groups}

The various components of the full physical wavefunction \eqn{Psiphyscomp} can now
be combined together using the results of the previous section and section 3.1.
After some algebra, we arrive finally at
\bea
& &\Psi^{(q)}_{\rm phys}{a\choose b}[\theta,K,\gamma_l;t]\nn\\& &
=\exp\left[i\oint\limits_{\partial\Sigma_p(t)}K-i\sum_{m,l=1}^{N_{p-1}}
I_{lm}^{(p-1)}\,\nu[\tilde\Sigma_{d-p+1}^{(l)},\partial\Sigma_p(0)]
\oint\limits_{\tilde\Sigma_{p-1}^{(m)}}K\right.
\nn\\& &~~~\left.+\,(-1)^{p(d-1)}\,i\oint\limits_
{\partial\Sigma_{d-p}(t)}\theta-(-1)^{p(d-1)}\,i\sum_{l,m=1}^{N_{p+1}}
I_{lm}^{(p+1)}\,\nu[\tilde\Sigma_{p+1}^{(l)},\partial\Sigma_{d-p}(0)]
\oint\limits_{\tilde\Sigma_{d-p-1}^{(m)}}\theta\right]
\nn\\& &~~~\times\,\exp\left\{-\frac{2\pi i}{\Omega_{d-1}k}\,Q_p
\,Q_{d-p}\,(-1)^{p(d-p)}\Bigl[\Phi_p(t)-\Phi_p(0)\Bigr]\right.\nn
\\& &~~~-\,(-1)^{p(d-p)}\,\frac{2\pi i}k\,Q_p\,Q_{d-p}\int\limits_0^tdt'~
\sum_{l,m=1}^{N_p}\Sigma_m^{(p)}(t')\,I^{(p)lm}\int\limits_0^{t'}dt''~
\Sigma_l^{(d-p)}(t'')\nn\\& &~~~-\,(-1)^{p(d-p)}\,i\,Q_{d-p}\sum_{m,n,r=1}^{N_p}
\int\limits_0^tdt'~\Sigma_m^{(d-p)}(t')\,I^{(p)mr}G_{rn}\,\gamma^n+
\frac k{4\pi}\sum_{l,r=1}^{N_p}\gamma^l\,G_{lr}\,\gamma^r\nn\\& &~~~\left.
-\,(-1)^{p(d-p)}\,\frac{\pi i}k\,Q_{d-p}^2\sum_{m,n,l,q=1}^{N_p}
\int\limits_0^tdt'~\Sigma_m^{(d-p)}(t')\,I^{(p)nm}\tau_{nl}
I^{(p)lq}\int\limits_0^tdt''~\Sigma_q^{(d-p)}(t'')\right\}\nn\\& &~~~
\times\,\Theta{\frac{a+q}{{\cal I}^{(p)}k_1k_2}\choose b}\left(
\frac{{\cal I}^{(p)}k_1}{2\pi}\,(-1)^{p(d-p)}\sum_{m=1}^{N_p}I_{ml}^{(p)}\,\gamma^m
\right.\nn\\& &~~~\left.\left.-\,(-1)^{p(d-p)}k_2\int\limits_0^tdt'~\left[Q_p
\,\Sigma_l^{(p)}(t')-Q_{d-p}\sum_{n,q=1}^{N_p}\tau_{ln}I^{(p)nq}\,
\Sigma_q^{(d-p)}(t')\right]\right|-k_1k_2{\cal I}^{(p)}\tau\right)
\nn\\& &
\label{totalwavefn}\eea
where $q^l=1,\dots,{\cal I}^{(p)}k_1k_2$, $l=1,\dots,N_p$, the $BF$ coefficient
$k$ is given by \eqn{coeffquant}, and the harmonic parts of the source degrees 
of freedom $\Sigma_l^{(p)}(t)$ are determined by the period integrals
\eqn{Dcomps} of the corresponding harmonic forms over the trajectories
$\Sigma_p(t)$. The external charge parameters in \eqn{totalwavefn} are in addition
constrained by the topological quantization condition \eqn{topquant}.
In this section we will study various aspects of the transformation
properties of the physical states \eqn{totalwavefn} in connection with 
the representation theory of the associated motion group. To set the
stage for this, we begin by describing some general aspects of motion groups.

\newsubsection{The Dahm Motion Group}

The Dahm motion group \cite{dahm,goldsmith} of a compact subspace
$\Sigma\subset\MD$ is the group of essentially different ways of 
continuously propagating $\Sigma$ in $\MD$ so that at the end of the motion,
$\Sigma$ returns to its original configuration in $\MD$. This topological
structure generalizes the Artin braid group \cite{braids}, whereby a braid is viewed as
a continuous one-parameter family of trajectories of $N$ distinct points
in the plane, where at each time $t_0$, the configuration is given by
the intersection of the braid at height $z=t_0$. Thus, the motion group has its
origins in the Artin braid group, and therefore the present topological
field theory yields the appropriate generalization of fractional statistics of
(extended) objects to {\it any} dimension. These applications will be 
discussed in the next section.

Let us start by recalling the definition of the braid group 
$B_N({\cal M}_2)$ of a connected Riemann surface ${\cal M}_2$ of genus $g$.
It can be constructed as the fundamental group
of the quantum mechanical configuration space $Q_N({\cal M}_2)$ for the
motion of $N$ identical particles on ${\cal M}_2$ with a hard-core 
repulsive interaction between them:
\beq
Q_N({\cal M}_2)=({\cal M}_2^N-\Delta_N)/S_N
\label{QNR2}\eeq
where $\Delta_N=\{(x,x,\dots,x)\,|\,x\in{\cal M}_2\}$ is the diagonal subspace of
${\cal M}_2^N$. Let $\vec z\in{\cal M}_2^N$ be some configuration of $N$ points in
${\cal M}_2$. A {\it motion} of $\vec z$ in ${\cal M}_2$ is a loop $\vec z(t)\in
Q_N({\cal M}_2)$, $t\in[0,1]$, based at $\vec z$. The {\it group of motions} of
$\vec z$ in ${\cal M}_2$ is then defined as the fundamental homotopy group
$\pi_1(Q_N({\cal M}_2);\vec z)$ of loops in $Q_N({\cal M}_2)$ based at $\vec z$. Since
the configuration space \eqn{QNR2} is a connected manifold ($\pi_0(Q_N({\cal M}_2))
=0$), one may prove that
\beq
\pi_1(Q_N({\cal M}_2);\vec z)=B_N({\cal M}_2)
\label{pi1BN}\eeq
The generators of $B_N({\cal M}_2)$ are given by the operators
$\sigma_n$, $n=1,\dots,N-1$, which braid the trajectories of particles $n$ and
$n+1$, along with the usual presentation of the Artin braid group of the plane
${\cal M}_2=\IR^2$ \cite{braids}.
In addition, there are $2g$ generators associated
with carrying a particle trajectory around each homology generator of
${\cal M}_2$ \cite{ladeg}.

It is precisely this homotopy definition of the braid group that generalizes
and gives the general notion of a motion group, whereby we replace the Riemann
surface ${\cal M}_2$
with an arbitrary $d$-manifold $\MD$ and the collection $\vec z$ of $N$ points
in ${\cal M}_2$ by any compact subspace $\Sigma\subset\MD$. We then define a
{\it motion} $f$ of $\Sigma$ in $\MD$ to be a path $f_t$, $t\in[0,1]$, of
homeomorphisms of $\MD$ with compact support such that $f_0=\id_{\MD}$ and
$f_1(\Sigma)=\Sigma$. A {\it stationary motion} of $\Sigma$ in $\MD$ is a
motion $f$ for which $f_t(\Sigma)=\Sigma~~\forall t\in[0,1]$. The product
$f\cdot g$ of two motions is the path
\beq
(f\cdot g)_t=\left\{\new{\begin{array}{cll}
g_{2t}&~~,~~&0\leq t\leq\mbox{$\frac12$}\\
f_{2(t-\frac12)}\circ g_1&~~,~~&\mbox{$\frac12$}\leq t\leq1\end{array}}
\right.
\label{motionprod}\eeq
while the inverse $f^{-1}$ of a motion $f$ is the path $f_{1-t}\circ
f_1^{-1}$. We say that two motions $f,g$ of $\Sigma$ in $\MD$ are equivalent,
$f\equiv g$, if $f^{-1}\cdot g$ is homotopic to a stationary motion. In
particular, stationary motions are equivalent to the trivial motion
$f_t=i_\Sigma~~\forall t\in[0,1]$, where $i_\Sigma:\Sigma\hookrightarrow\MD$
is the canonical inclusion. It may then be shown that the corresponding set of
equivalence classes of motions of $\Sigma$ in $\MD$, with the multiplication
induced by that in \eqn{motionprod}, forms a group $\orbit_\Sigma(\MD)$ which 
is called
the {\it Dahm motion group} of $\Sigma$ in $\MD$ \cite{dahm,goldsmith}.

To understand what this group represents in terms of configuration space
homotopy, let $e(\MD,\Sigma)$ denote the space of embeddings of $\Sigma$
in $\MD$ and $h(\MD)$ the space of homeomorphisms of $\MD$ of compact
support, both equipped with the compact-open topology. Let $h(\MD,\Sigma)
\subset h(\MD)$ be the subspace of homeomorphisms which leave $\Sigma$
fixed. Note that both $h(\MD)$ and $h(\MD,\Sigma)$ are topological groups,
so that one may define the fundamental relative homotopy group 
$\pi_1(h(\MD),h(\MD,\Sigma);\id_{\MD})$ as the set of homotopy classes of
paths in $h(\MD)$ which begin at $\id_{\MD}\in h(\MD,\Sigma)$ and end in
$h(\MD,\Sigma)$, and with multiplication induced by that of $h(\MD)$. Then
by definition it follows that
\beq
\orbit_\Sigma(\MD)=\pi_1(h(\MD)\,,\,h(\MD,\Sigma)\,;\,\id_{\MD})
\label{motiongpdef}\eeq
Note that the analog of the quantum configuration space \eqn{QNR2} in the
situation at hand is the quotient space $e(\MD,\Sigma)/\sim$, where 
$f\sim f'$ if $f(\Sigma)=f'(\Sigma)$, but the topology of this space is
unmanageable. However, it is straightforward to show \cite{goldsmith} that if $f,g$ are
motions of $\Sigma$ in $\MD$, then $f\equiv g$ if and only if $f$ is 
homotopic to a motion $f'$ of $\Sigma$ in $\MD$ with 
$f_t'(\Sigma)=g_t(\Sigma)~~\forall t\in[0,1]$. With this property, it is
evident that the above definition of a motion concides with the notion of a
loop in a quantum configuration space.

Let us now consider some basic examples of motion groups. First, we note
that any $k$-isotopy of $N$ distinct points in a manifold $\MD$ extends to
a $k$-isotopy of all of $\MD$. From this fact one may prove that, if
$P_N=\{x_1,\dots,x_N\}\subset\MD$ is a collection of $N$ distinct points of
$\MD$, then the restriction map
\beq
(h(\MD)\,,\,h(\MD,P_N)\,,\,\id_{\MD})\stackrel{\rho}{\longrightarrow}
(e(\MD,P_N)\,,\,e(P_N,P_N)\,,\,\id_{P_N})
\label{restrmap}\eeq
induces isomorphisms
\beq
\rho^*:\,\pi_n(h(\MD)\,,\,h(\MD,P_N)\,;\,\id_{\MD})\stackrel{\approx}
{\longrightarrow}\pi_n(e(\MD,P_N)\,,\,e(P_N,P_N)\,;\,\id_{P_N})
\label{restrisos}\eeq
for all $n\geq0$ \cite{goldsmith}.
It follows that the group of motions of a point $x\in\MD$
coincides with the fundamental group
\beq
\orbit_x(\MD)=\pi_1(\MD,x)
\label{motionp}\eeq
and, if $\MD$ is connected, the motion group of $P_N$ is the braid group
\beq
\orbit_{P_N}(\MD)=B_N(\MD)
\label{motionbraid}\eeq
Notice that, if $\dim\MD>2$, then \cite{dahm}
\beq
\orbit_{P_N}(\MD)\cong\bigoplus_{n=1}^N\pi_1(\MD;x_n)
\label{motion2}\eeq
from which it follows that the braiding phenomenon disappears in manifolds
of dimension larger than 2. A more interesting example is provided by the
Dahm group of motions of $N$ unlinked and unknotted circles in $\IR^3$. The
generators are the motions which flip a circle, exchange two circles, and move
one circle through another \cite{goldsmith}. We shall return to this example in
the next section.
For a
description of the Dahm group of a certain class of non-trivial links in
${\bf S}^3$, see \cite{goldsmith1}.

The present $BF$ field theory approach that we are ultimately interested in
actually focuses on {\it two} disjoint, compact submanifolds 
$\Sigma=\partial\Sigma_p(0)$ and $\Sigma'=\partial\Sigma_{d-p}(0)$ which
lead to motions $\Sigma_p(t)$ and $\Sigma_{d-p}(t)$ of dual
dimension in $\MD$.
For this, we need a slight generalization of the Dahm motion group above
\cite{goldsmith}.
Let $h(\MD,\Sigma,\Sigma')$ be the subspace of $h(\MD)$ of homeomorphisms
which leave fixed {\it both} of the submanifolds $\Sigma$ and $\Sigma'$. Then 
the motion group of the pair $(\Sigma,\Sigma')$ in $\MD$ is defined to be
the relative fundamental homotopy group:
\beq
\orbit_{\Sigma,\Sigma'}(\MD)=\pi_1(h(\MD)\,,\,h(\MD,\Sigma,\Sigma')\,;\,
\id_{\MD})
\label{motiongppair}\eeq
Note that this definition differs from that of \eqn{motiongpdef} with
$\Sigma$ replaced by $\Sigma\amalg\Sigma'$, since this latter group uses 
motions of $\Sigma$ and $\Sigma'$ which return to their original value
only modulo permutation of the two subspaces $\Sigma,\Sigma'$. The former group,
on the other hand, is the one that is required when the statistics of the
branes corresponding to $\Sigma,\Sigma'$ are non-identical, as is the case
in the canonical formulation of $BF$ field theory which utilizes brane
source couplings to two independent gauge fields.
In fact, the relation between the motion group \eqn{motiongppair} and the
Dahm groups
\eqn{motiongpdef} is given by the following theorem
\cite{goldsmith}. Let $\varepsilon:
\orbit_{\Sigma'}(\MD-\Sigma)\to\orbit_{\Sigma,\Sigma'}(\MD)$ be the group
homomorphism induced by the map
\beq
(h(\MD-\Sigma)\,,\,h(\MD-\Sigma,\Sigma')\,,\,\id_{\MD-\Sigma})\longrightarrow
(h(\MD)\,,\,h(\MD,\Sigma,\Sigma')\,,\,\id_{\MD})
\label{epindmap}\eeq
which sends each homeomorphism 
$f\in h(\MD-\Sigma)$ to its extension $\varepsilon(f)
\in h(\MD)$ with $\varepsilon(f)|_\Sigma=\id_\Sigma$. Let $\varpi:
\orbit_{\Sigma,\Sigma'}(\MD)\to\orbit_\Sigma(\MD)$ be the group
homomorphism induced by the map
\beq
(h(\MD)\,,\,h(\MD,\Sigma,\Sigma')\,,\,\id_{\MD})\longrightarrow
(h(\MD)\,,\,h(\MD,\Sigma)\,,\,\id_{\MD})
\label{piindmap}\eeq
which sends each $f\in h(\MD,\Sigma,\Sigma')$ to $f\in h(\MD,\Sigma)$. Then
the sequence of groups
\beq
\orbit_{\Sigma'}(\MD-\Sigma)\stackrel{\varepsilon}{\longrightarrow}
\orbit_{\Sigma,\Sigma'}(\MD)\stackrel{\varpi}{\longrightarrow}
\orbit_\Sigma(\MD)
\label{motionexact}\eeq
is exact.

We close this subsection with a final useful computational property of
the Dahm motion groups. Namely, there is a map, called the {\it Dahm
homomorphism} \cite{dahm,goldsmith}, which is a homomorphism
\beq
{\cal D}:\,\orbit_\Sigma(\MD)\longrightarrow{\rm Aut}(\pi_1(\MD-\Sigma))
\label{dahmhomo}\eeq
from the group of motions of $\Sigma$ in $\MD$ to the automorphism group
of $\pi_1(\MD-\Sigma)$ induced at the end of a given motion. This yields 
another presentation of the motion group which may be thought of as the
fundamental homotopy group of an appropriate quantum configuration space.
For some other
computational aspects of the motion groups, see \cite{goldsmith}.
In the following we will derive a class of representations of the motion
group $\orbit_{\Sigma,\Sigma'}(\MD)$ which illustrates some new general
aspects of these topological groups, and thereby extends the generally
unprobed theory of motion groups. This will present a highly non-trivial
application and thereby demonstrate the usefulness of the present 
topological field theory approach.

\newsubsection{Abelian Holonomy Representations}

Let us now examine the various transformation properties of the wavefunctions
\eqn{totalwavefn} and describe the ensuing representations of the motion group.
Consider an adiabatic motion of the hypersurface $\Sigma_p(t)$ about $\Sigma
_{d-p}(t)$. First we consider the homologically trivial motions. If one of the
hypersurfaces $\Sigma_p(t)$ or $\Sigma_{d-p}(t)$ traces out a contractible
volume as it moves, then the topological current integrals in the full
wavefunction \eqn{totalwavefn} vanish (c.f. eq. \eqn{Dcomps}). These currents
therefore contribute nothing to the physical states under these types of
motions. The solid angle function $\Phi_p(t)-\Phi_p(0)$, on the other hand,
has delta-function singularities and thereby contributes whenever the hypersurfaces
link each other. The change in $\Phi_p$ whenever such a linking occurs is
$\Omega_{d-1}$, but this function is nevertheless independent of the choice of
representative of the topological classes of the source trajectories, as 
shown in section 4.2. Thus under such an adiabatical linking, the
wavefunctions \eqn{totalwavefn} acquire the phase
\beq
(\sigma_p)^2=\e^{-\frac{2\pi i}k\,(-1)^{p(d-p)}Q_pQ_{d-p}}
\label{sigmapphase}\eeq

Now let us examine the case of a homologically non-trivial motion. Consider
the source motion whereby $\Sigma_p(t)$ is fixed in time and $\Sigma_{d-p}(t)$
winds $w_l^{d-p}$ times, in a time span $t_0$, around the $l$-th homology
$d-p$-cycle of $\MD$, and then afterwards $\Sigma_{d-p}(t)$ is fixed and 
$\Sigma_p(t)$ winds $w_l^p$ times, up to some time $t>t_0$, around the 
$l$-th homology $p$-th cycle of $\MD$. According to \eqn{Dcomps}, this motion can
be summarized by the following equations:
\bea
\int\limits_0^{t_0}dt'~\Sigma_l^{(d-p)}(t')=w_l^{d-p}~~~~~~&,&~~~~~~
\int\limits_0^{t_0}dt'~\Sigma_l^{(p)}(t')=0\nn\\\int\limits_{t_0}^tdt'~
\Sigma_l^{(d-p)}(t')=0~~~~~~&,&~~~~~~\int\limits_{t_0}^tdt'~\Sigma_l
^{(p)}(t')=w_l^p
\label{topintshommotion}\eea
The holonomies arising from possible linkings of these motions are taken
into account by the solid angle function and constitute the phase operators
\eqn{sigmapphase} for the motion group. The remaining part of the periodic
motion is readily found to change the wavefunctions \eqn{totalwavefn} according to
\beq
\Psi_{\rm phys}^{(q)}[\theta,K,\gamma;t]\longmapsto\sum_{q'}\left[
M(w^p,w^{d-p})\right]_{qq'}\,\Psi_{\rm phys}^{(q')}[\theta,K,\gamma;0]
\label{Psihomchange}\eeq
where the unitary matrices
\bea
\left[M(w^p,w^{d-p})\right]_{qq'}&=&\prod_{l=1}^{N_p}\exp\left[-
\frac{2\pi i}{k_1{\cal I}^{(p)}}\,(-1)^{p(d-p)}
\left(Q_pQ_{d-p}k_2\sum_{m=1}^{N_p}w_m^pI^{(p)lm}w_l^{d-p}\right.\right.
\nn\\& &\left.\left.+\,Q_{d-p}
\sum_{m=1}^{N_p}b_mI^{(p)ml}w_l^{d-p}-Q_pw_l^pq^l-Q_pw_l^pa^l\right)
\right]\nn\\& &\times\,\delta_{q^r-k_2\sum_sI^{(p)rs}w_s^{d-p}\,,\,q'^r}
\label{Mdual}\eea
generate a $(k_1)^{N_p}$-dimensional projective representation of the group
$\Gamma\oplus\Gamma^*$ of large gauge transformations. Their products also
determine a global $U(1)\times U(1)$ two-cocycle algebra:
\bea
M(w^p,w^{d-p})M(v^p,v^{d-p})&=&\prod_{l,m=1}^{N_p}\e^{-\frac{2\pi i}k\,(-1)
^{p(d-p)}Q_pQ_{d-p}\left(w_l^pv_m^{d-p}-v_l^pw_m^{d-p}\right)I^{(p)lm}}
\nn\\& &\times\,M(v^p,v^{d-p})M(w^p,w^{d-p})
\label{dualclockalg}\eea
which may be viewed as {\it dual} to the algebra \eqn{clockalg} of the winding
translation generators, in that the integers $k_1$ and $k_2$ are interchanged
and the intersection matrix $I_{lm}^{(p)}$ is replaced by its inverse through
the combination $Q_pQ_{d-p}I^{(p)ml}$ (The appearence of the charges here owes to
the fact that the algebra \eqn{dualclockalg} comes from the windings of the
sources, whereas the algebra \eqn{clockalg} comes about from the windings
of the gauge fields themselves).

To describe the appropriate motion group representation, we define $w_k^{p(l)}=\delta
_k^l$ and introduce the unitary operators
\beq
\eta_p^{(l)}=M(w^{p(l)},0)~~~~~~,~~~~~~\mu_p^{(m)}=M(0,w^{d-p(m)})
\label{etamudef}\eeq
for each $l,m=1,\dots,N_p$. Then, together with the phase operators $\sigma_p
\id_{(k_1)^{N_p}}$, these operators generate the following $(k_1)^{N_p}$
dimensional representation of the pertinent motion group:
\bea
\left[\eta_p^{(l)}\,,\,\eta_p^{(m)}\right]&=&\left[\mu_p^{(l)}\,,\,\mu_p
^{(m)}\right]~=~0\label{M1a}\\\left[\sigma_p\,,\,\eta_p^{(l)}\right]
&=&\left[\sigma_p\,,\,\mu_p^{(l)}\right]~=~0\label{M1}\\\eta_p^{(l)}
\mu_p^{(m)}&=&(\sigma_p)^{2I^{(p)lm}}\,\mu_p^{(m)}\eta_p^{(l)}
\label{M2}\eea
and, according to \eqn{topquant}, the topology of the manifold $\MD$
imposes the following global constraint on these generators:
\beq
1=\prod_{r,l=1}^{N_{p-1}}\,\prod_{m,n=1}^{N_{p+1}}\Bigl(\sigma_p\Bigr)
^{2\nu[\tilde\Sigma_{p+1}^{(m)},\partial\Sigma_{d-p}(0)]\,\nu[\tilde\Sigma
_{d-p+1}^{(l)},\partial\Sigma_p(0)]\left(I_{mn}^{(p+1)}+I_{rl}^{(p-1)}
\right)}
\label{M3}\eeq
The collection of $2N_p+1$ unitary operators $\{\sigma_p,\eta_p^{(l)},\mu
_p^{(m)}\}$ with the relations \eqn{M1a}--\eqn{M3} constitute a subset of
the full set of generators of a $(k_1)^{N_p}$ dimensional representation of
the motion group $\orbit_{\partial\Sigma_p(0),\partial\Sigma_{d-p}(0)}(\MD)$.
Presumably there are more generators and relations 
for this group (see the next section), but
due to the abelian nature of the present formalism such operators are
represented trivially on the physical Hilbert space of the $BF$ field theory.
Notice that the constraint \eqn{M3} is represented in terms of the intersection
matrices $I^{(p-1)}$ and $I^{(p+1)}=-I^{(d-p-1)}$ (by Poincar\'e-Hodge duality)
which arise from the initial $p-1$-brane and $d-p-1$-brane configurations
$\partial\Sigma_p(0)$ and $\partial\Sigma_{d-p}(0)$ used to define the
appropriate motion group (and which come about from the relevant \v{C}ech
cohomology groups).

The global constraint \eqn{M3} comes about from the fact that there is always 
a trajectory $\Sigma_p(t)$ which encircles $\Sigma_{d-p}(t)$ and traces the 
homology generators of $\MD$ in such a way that it forms a trivial motion,
i.e. one that is equivalent to a stationary motion of $\partial\Sigma_p(0)$
in $\MD$. As a simple example of this constraint, consider the motion in $\MD$
whereby, initially at time $t=0$, the $p-1$-brane intersects {\it only}
with the $l_0$-th homology $d-p+1$-cycle of $\MD$ exactly once, and likewise
for the $d-p-1$-brane with the $m_0$-th homology $p+1$-cycle. 
Then the
global restriction \eqn{M3} simplifies to the form
\beq
\Bigl(\sigma_p\Bigr)^{2\sum_nI_{m_0n}^{(p+1)}+2\sum_rI_{rl_0}^{(p-1)}}=1
\label{M3simpl}\eeq
This relation is a fundamental constraint that must be met the external 
charges $Q_p,Q_{d-p}$ of the quantum field
theory in order to yield a well-defined motion
group representation.
Generally, the relations \eqn{M1a}--\eqn{M2} between the linking operator $\sigma_p$
and the generators of homologically non-trivial motions $\eta_p^{(l)},
\mu_p^{(m)}$
reflect the non-trivial relationships that exist between
motions around the various cycles of $\MD$. The relation \eqn{M2} is very
natural, since it tells us that the operations of moving $\Sigma_p$ and $\Sigma_{d-p}$
around the pertinent cycles commute {\it only} when these cycles do not
intersect. Otherwise, they differ by a holonomy factor that depends precisely on
the intersection number of the cycles and represents the number of linking
operations required to unravel the motion to a stationary one. Together with
the exact sequence \eqn{motionexact}, these relationships may determine at least
a large portion of the full motion group for a wide class of submanifold
embeddings in terms of their individual motions in $\MD$. These relationships
thereby reflect a highly non-trivial application of the present topological field
theory to the theory of motion groups. In the next section we shall describe
briefly how the present model may be modified so as to potentially produce
the full set of generators and motions of $\orbit_{\partial\Sigma_p(0),
\partial\Sigma_{d-p}(0)}(\MD)$.

\newsection{Applications}

In this final section we will briefly describe some examples and applications of
the formalism above. We will also mention possible generalizations which could
probe deeper into the structure of motion groups on topologically non-trivial 
manifolds.

\newsubsection{The Braid Group}

For our first example, we illustrate how the well-known holonomy representations
of the braid group \cite{djt}--\cite{bs}
appear within our more general formalism. We set $d=2$,
$p=1$ and $N_1=2g$, where $g$ is the genus of a compact Riemann surface 
${\cal M}_2$. For any $g>0$, we may view ${\cal M}_2$ as the connected sum
$({\bf T}^2)^{\#g}$ of two-tori and hence
as an embedded submanifold of $\IR^3$. The canonical
basis of $H_1({\cal M}_2,\zed)\cong\zed^{2g}$ is defined by the $2g$ generators
$a^l$ and $b_m$, $l,m=1,\dots,g$, where $a^l$ corresponds to the class of the
outer cycle of ${\bf T}^2$ in the $l$-th component of the connected sum
$({\bf T}^2)^{\#g}$ while $b_m$ corresponds to the class of the inner cycle of
${\bf T}^2$ in the $m$-th component. The intersection indices of these one-cycles are
given by
\bea
\nu[a^l,a^k]&=&\nu[b_l,b_k]~=~0\nn\\\nu[a^l,b_k]&=&-\nu[b_k,a^l]~=~\delta_k^l
\label{intindsab}\eea
and the corresponding intersection matrix is
\beq
I^{(1)kl}=\pmatrix{0&\id_g\cr-\id_g&0\cr}
\label{I1kl}\eeq
with $k,l=1,\dots,2g$.
Furthermore, we have $I^{(0)}=-I^{(2)}=\pm1$, where the sign is chosen with 
respect to a given orientation of the surface ${\cal M}_2$ when viewed as an
embedded submanifold of $\IR^3$.

In the present case, we denote the generators $\eta_1^{(l)}$ (and $\mu_1^{(l)}$)
by $\alpha^l$ for $l=1,\dots,g$ and by $\beta_{m=2g+1-l}$ for 
$l=g+1,\dots,2g$. The operator which represents the braiding of the trajectories
of two particles of charges $Q_1$ and $Q_2$ is
\beq
\sigma=\e^{\frac{\pi i}k\,Q_1Q_2}
\label{braidop}\eeq
Then, according to the relations \eqn{M1a}--\eqn{M2}, these operators have the
following presentation which represents various equivalent braids:
\bea
\left[\sigma\,,\,\alpha^l\right]&=&\Bigl[\sigma\,,\,\beta_l\Bigr]~=~0\nn\\
\left[\alpha^l\,,\,\alpha^m\right]&=&\Bigl[\beta_l\,,\,\beta_m\Bigr]~=~0\nn\\
\left[\alpha^l\,,\,\beta_m\right]&=&0~~~~{\rm for}~~l\neq m\nn\\
\alpha^l\,\beta_l&=&\sigma^2\,\beta_l\,\alpha^l
\label{braidrels}\eea
for each $l,m=1,\dots,g$. However, the global constraint \eqn{M3} for this
particular braid group representation is an identity. The topology of the
Riemann surface ${\cal M}_2$ in the present case does {\it not} affect the 
linking operator $\sigma$, because the $BF$ field theory representation in
effect generates an abelian holonomy representation of the {\it unpermuted}
braid group of ${\cal M}_2$, i.e. that associated with the quantum mechanical
configuration space of a system of {\it non-identical} particles. For a system
of $N$ identical particles, the corresponding braid group representation would
have to satisfy the additional global constraint $\sigma^{2(N+g-1)}=1$
for a closed manifold \cite{ladeg,bs}. Normally, such a constraint would come geometrically
in part from a
framing of the corresponding three-dimensional manifold which is required to
regulate the self-linkings of the particle trajectories \cite{tze}. Here the linking
numbers induced by the $BF$ field theory contain no such ambiguous self-linking
terms. In effect, the present holonomy factors induce a representation of
the two colour braid group \cite{rolfsen} generated by exchanging ribbon-like configurations. 
The ribbons can themselves twist, leading to intrinsic spin phases which
cancel exactly with the statistical exchange phases as a result of the 
spin-statistics theorem (This may be checked explicitly by computing the
action of the energy-momentum tensor on the physical states \eqn{totalwavefn}).
Note that the argument of the solid angle function
\eqn{Phidef} defines an eigenfunction expansion of the prime form of the 
Riemann surface ${\cal M}_2$ which in turn produces the appropriate generalization
of the usual multi-valued angle function of the plane \cite{djt}--\cite{bs}. 

\newsubsection{Quantum Exchange Statistics of Extended Objects}

The generic properties of the braid group representations described above can be
generalized to any spatial manifold $\MD$ of even dimension $d=2p$. This in
turn provides a field theoretical model which generalizes the phenomenon of
fractional statistics of quantum mechanical point particles in two dimensions
to that of non-identical $p-1$-branes in $2p$-dimensions (strings in
four spatial dimensions, membranes in six spatial dimensions, etc.). Note that
to describe a system of $N>2$ non-interacting
objects, one considers the worldvolumes to be
disjoint unions $\Sigma_p=\coprod_{n=1}^N\Sigma_p^{(n)}$ and writes the
corresponding deRham current of $\Sigma_p$ as a sum over those of the
$\Sigma_p^{(n)}$. In the present case, we again have that $I^{(p-1)}=-I^{(p+1)}$
(by Poincar\'e-Hodge duality) and so the global constraint \eqn{M3} again
simplifies considerably. In particular, if the initial configurations at
time $t=0$ of both $p-1$-branes intersect the homology $p+1$-cycles of
${\cal M}_{2p}$ in the same way, then this global constraint is once again
an identity and there are no further constraints on the linking operators
$\sigma_p$ of the holonomical motion group representation. This feature
again owes to the fact that we obtain a representation of the {\it unpermuted},
two-colour motion group. As in the case of particles, we attribute this global
cancellation as being due to an induced spin of the extended motions which
cancels the holonomy factors. In particular, we may deduce from this
cancellation that the standard spin-statistics theorem holds for such
configurations of extended objects (in contrast to the generic case \cite{bmosss}).

In the special case where the spatial manifold is flat infinite Euclidean
space $\MD=\IR^d$, the constructions of the previous sections can be made somewhat
more explicit (essentially because there are no harmonic zero modes in this
case). In particular, it is known \cite{goldsmith} that in this case the motion
groups count the connected components of the space of orientation-preserving
homeomorphisms of $\IR^d$ which preserve $\Sigma$,
\beq
\orbit_\Sigma(\IR^d)\cong\pi_0(h^+(\IR^d,\Sigma))
\label{motionreald}\eeq
Furthermore, the Euclidean Green's function of the scalar Laplacian for $p>1$
is given by
\beq
\left(x\left|\nabla_0^{-2}\right|y\right)=-\frac1{\Omega_{2p-1}|x-y|^{2(p-1)}}
\label{Euclgreenfn}\eeq
Then, using \eqn{derhamcurrent}, \eqn{Drhocomp} and \eqn{tildePsisoln}, we arrive
at an explicit expression for the (symmetrized)
$2p-1$-dimensional solid angle formed between
two $p-1$-brane configurations at time $t$:
\bea
\Phi_p(t)&=&\frac{2(p-1)}{\Omega_{2p-1}}\int\limits_0^tdt'~\int d^{p-1}\sigma~
\int d^{p-1}\sigma'~\epsilon_{0i_1\cdots i_p}\nn\\& &\times\left[\prod_{k=1}^{p-1}
\frac{\partial X_p^{i_k}(t',\sigma)}{\partial\sigma^k}\,\frac{\partial
X_p'^{i_{p+1}}(t',\sigma')}{\partial t'}\,\prod_{l=p+2}^{2p}\frac{\partial
X_p'^{i_l}(t',\sigma')}{\partial\sigma'^l}\right.\nn\\& &\left.-\prod_
{k=1}^{p-1}\frac{\partial X_p'^{i_k}(t',\sigma')}{\partial\sigma'^k}\,
\frac{\partial X_p^{i_{p+1}}(t',\sigma)}{\partial t'}\,\prod_{l=p+2}^{2p}
\frac{\partial X_p^{i_l}(t',\sigma)}{\partial\sigma^l}\right]\,
\frac{\left(X_p(t',\sigma)-X_p'(t',\sigma')\right)^{i_p}}{\left|X_p(t',\sigma)-
X_p'(t',\sigma')\right|^{2p}}\nn\\& &
\label{Phipevend}\eea
The present topological field theory formalism therefore produces very explicit
higher-dimensional generalizations of the standard multi-valued angle functions
which have been extensively studied and utilized in the physics of planar
systems \cite{djt}--\cite{bs}. 
Eq. \eqn{Phipevend} generalizes the standard expression for the adiabatic
limit of the Gauss linking number in three-dimensions for two curves ($p=1$)
\cite{tze}. In
this way the function \eqn{Phipevend} may in fact be thought of as giving the 
appropriate higher-rank generalization of the electromagnetic Faraday law, but in
a way that avoids the cumbersome self-linking number terms that arise in the standard
two dimensional formulations \cite{djt}--\cite{bs},\cite{tze}.

\newsubsection{Quantum Exchange Statistics in Odd Dimensions}

Fractional statistics of identical extended objects in odd dimensional spaces may be
attained by modifying the topological class of one of the deRham currents appearing
in the source-coupled $BF$ action. An important example is the case of the motion
of $N$ non-interacting identical strings in flat Euclidean three-space $\IR^3$.
The relevant motion group $\orbit_{\Sigma^{(N)}}(\IR^3)$ in this case is constructed from
the submanifold $\Sigma^{(N)}=\coprod_{n=1}^NC_n$ which is a collection of $N$ unknotted
and unlinked circles $C_n={\bf S}^1$ in $\IR^3$. Then
\beq
\pi_1(\IR^3-\Sigma^{(N)})
\cong\langle\ell_1,\dots,\ell_N\rangle
\label{freegp}\eeq
is the free group on $N$ generators
$\ell_n$, $n=1,\dots,N$. The generating automorphisms for the Dahm subgroup
${\cal D}(\orbit_{\Sigma^{(N)}}(\IR^3))$ of ${\rm Aut}(\langle\ell_1,\dots,\ell_N\rangle)$
are then $\tau_n$, $\sigma_n$ and $\rho_{nm}$ \cite{goldsmith}, where
\bea
\tau_n(\ell_k)&=&\left\{\new{\begin{array}{cll}
(\ell_n)^{-1}~~~~&,&~~k=n\\\ell_k~~~~&,&~~k\neq n\end{array}}\right.
\nn\\\sigma_n(\ell_k)&=&
\left\{\new{\begin{array}{cll}
\ell_{n+1}~~~~&,&~~k=n\\\ell_n~~~~&,&~~k=n+1\\\ell_k~~~~&,&~~k\neq n,n+1
\end{array}}\right.\nn\\\rho_{nm}(\ell_k)&=&\left\{\new{\begin{array}{cll}
\ell_m\,\ell_n\,(\ell_m)^{-1}~~~~&,&~~k=n\\\ell_k~~~~&,&~~k\neq n\end{array}}
\right.
\label{autgens}\eea
These automorphisms correspond, respectively, to a rotation through angle $\pi$
of the $n$-th circle $C_n$ about its diameter, to interchanging the $n$-th and
$n+1$-th circles, and to transporting the $n$-th circle through the $m$-th circle.
Some of the relations of this motion group can also be thereby deduced to be
\cite{abks}:
\bea
(\tau_n)^2&=&\id\nn\\\tau_n\,\rho_{n,n+1}\,(\tau_n)^{-1}&=&\rho_{n,n+1}\nn\\
(\sigma_n)^2&=&\id\nn\\\sigma_n\,
\sigma_m&=&\sigma_m\,\sigma_n~~~~{\rm for}~~|m-n|\geq2
\nn\\\sigma_n\,\sigma_{n+1}\,\sigma_n&=&\sigma_{n+1}
\,\sigma_n\,\sigma_{n+1}\nn\\
(\tau_n\,\sigma_n)^4&=&(\tau_n\,\rho_{n,n+1})^2~=~\id
\label{autrels}\eea

We shall now construct the appropriate topological $BF$ field theory to describe this
group \cite{abks}.
We consider the usual $BF$ action \eqn{canaction} for $d=3,p=2$ with the
standard coupling of the two-form field $B$ to the total deRham current of the
string worldsheets $\Sigma_2=\coprod_{n=1}^N\Sigma_2^{(n)}$ and with
\beq
\Delta_{\Sigma_2}=\sum_{n=1}^N\phi_n\,\Delta_{\Sigma_2^{(n)}}
\label{derhamtotal}\eeq
where $\phi_n$ is the flux of string $n$ (Here we set the overall
charge parameters $Q_1,Q_2$ equal to 1). The crucial modification is in the 
coupling $A\wedge\star{\cal J}$
of the one-form field $A$, where
the new current $\cal J$ is represented by the vector field \cite{balss}
\beq
{\cal J}^\mu(x)=\sum_{n=1}^N\int d^2\sigma~\delta^{(4)}(x-X_n(\sigma))\,
\frac{\partial X_n^\mu(\sigma)}{\partial\sigma^\alpha}\,J_n^\alpha(\sigma)
\label{calJdef}\eeq
which is defined in terms of the conserved worldsheet current
\beq
J_n^\alpha(\sigma)=\epsilon^{\alpha\beta}\,\frac{\partial\varphi_n(\sigma)}
{\partial\sigma^\beta}
\label{worldsheetcurrent}\eeq
Here $X_n:\Sigma_2^{(n)}\to\IR^3$ is the worldsheet embedding of string $n$ and
$\varphi_n(\sigma)$ is some continuous function on $\Sigma_2^{(n)}$. We assume that
$d\varphi_n\in\Omega^1(\Sigma_2^{(n)})$ is a globally defined differential
one-form on the string worldsheet, but that the function $\varphi_n$ itself
is multi-valued. Performing a canonical split of the coordinates of the surface
$\Sigma_2^{(n)}$ with $\sigma^2\in[0,1]$ parametrizing the loop of the closed
string, we see that the current \eqn{worldsheetcurrent} induces a non-zero $U(1)$
charge on the worldsheet of string $n$:
\beq
q_n=\int\limits_0^1d\sigma^2~J_n^1(\sigma^1,\sigma^2)=\varphi_n(\sigma^1,1)-
\varphi_n(\sigma^1,0)
\label{inducedcharge}\eeq
This charge is a constant of the motion because both currents \eqn{calJdef} and
\eqn{worldsheetcurrent} define closed differential forms on $\IR^3$
and $\Sigma_2^{(n)}$,
respectively. The current \eqn{calJdef} can thereby be thought of
as a smeared particle current which serves as a smoothed-out induced deRham
current and which induces an electric charge on the string worldsheets.

The worldsheet scalar fields $\varphi_n(\sigma)$ can be regarded as dynamical degrees of
freedom in the field theory, in which case they are associated with the 
reparametrization invariances of the string surfaces. Fixing these functions to
some prescribed form, with $d\varphi_n$ in the cohomology classes appropriate to
the charges \eqn{inducedcharge}, ruins the invariance of the strings under 
diffeomorphisms of ${\bf S}^1$. Nevertheless, we will set $\varphi_n(\sigma^1,
\sigma^2)=q_n\sigma^2$ and calculate the resulting holonomies that arise in the
modified wavefunctions of the quantum field theory. Because the current
\eqn{calJdef} is a closed one-form, the relevant phase factor that appears in
the wavefunctions can be calculated in the same way as before, by using the 
Hodge decompositions for \eqn{calJdef} analogous to those of the usual (singular)
deRham currents. Following the steps which led to \eqn{Phipevend} using the 
Euclidean Green's function for the three-dimensional scalar Laplacian operator, this
straightforward calculation produces the holonomy function
\bea
\Phi_{2,2}(t)&=&\sum_{n,m=1}^N\frac{q_n\phi_m}{4\pi}\,\int\limits_0^tdt'~\int
\limits_0^1d\sigma~\int\limits_0^1d\sigma'~\epsilon_{0ijk}\left[\frac{\partial
X_n^i(t',\sigma)}{\partial t'}\,\frac{\partial X_m^k(t',\sigma')}{\partial
\sigma'}\right.\nn\\& &\left.-\frac{\partial X_m^i(t',\sigma')}{\partial t'}
\,\frac{\partial X_m^k(t',\sigma')}{\partial\sigma'}\right]\,
\frac{\left(X_n(t',\sigma)-X_m(t',\sigma')\right)^j}{\left|X_n(t',\sigma)-
X_m(t',\sigma')\right|^3}
\label{Phi22}\eea

In this abelian holonomy representation, the images of the flip operators
$\tau_n$ and the exchange generators $\sigma_n$ are all trivial on the Hilbert
space (see \eqn{autgens}). This owes to the property that the strings have
no abelian linking in three-dimensions, and also that the function \eqn{Phi22}
yields no representation of the ``self-interactions" of a given string
configuration. The slide operators $\rho_{nm}$, on the other hand, produce
non-trivial quantum phases in the wavefunctions under the adiabatic transport
of string $n$ through the loop of string $m$. The resulting one-dimensional,
unitary holonomy representation is therefore given by
\bea
\tau_n&=&\sigma_n~=~1\nn\\\rho_{nm}&=&\e^{\frac{\pi i}k\,q_n\phi_m}
\label{linkrep}\eea
The adiabatic holonomy in \eqn{linkrep} arises from the fact that electric
charge and flux can link in three-dimensions, so that the sliding operation
has the same effect as adiabatically transporting a charge $q_n$ around a
flux $\phi_m$ \cite{bss}. 
Whether or not this model, with a particular fixed configuration for the 
worldsheet scalar fields $\varphi_n(\sigma)$, leads to a sensible Hilbert space
representation is a point which deserves further investigation. In any case, this
simple example shows the possibilities that exist for constructing representations
of {\it arbitrary} motion groups for any dimensionalities of the manifolds involved.
It would also be interesting to analyse global aspects of these sorts of couplings
to $BF$ gauge fields, along the lines developed in earlier sections of this
paper. This construction would then yield the extra (multi-dimensional) generators 
and relations of the motion
group 
which arise due to homological effects, and also compute the curved space version 
of the holonomy function \eqn{Phi22}. A holonomy function for strings in 
generic three-dimensional spatial manifolds has been derived in \cite{me} based
on a marginal deformation of the canonical $BF$ field theory.

The representations obtained in this paper are all abelian (although
multi-dimensional when the space contains non-contractible cycles) and as such
lead to very simple representations of the generators and relations of
the motion group. For this reason they do not completely probe the 
algebraic structures of the motion group, although they do provide geometrical
and field theoretical origins for various aspects of it. More
interesting representations may be attainable using non-abelian $BF$ theories,
whereby the increase in colour symmetry is expected to give rise to richer
invariants of the embedded submanifolds and of the spatial manifolds themselves.
The holonomy operators in four-dimensional non-abelian $BF$-theory have 
been studied recently in \cite{ccr} where it was shown that surface
observables yield possibly new invariants of immersed surfaces in 
four-manifolds. A non-abelian version of the $BF$ model described in this subsection
is analysed briefly in \cite{abks}. It would also be interesting to
incorporate ``interactions" of various extended objects into the framework of
this paper. This would lead to a quantum field theoretical description of, for
example, the motion groups associated with non-trivial knots and links immersed
in $\IR^3$ such as those studied in \cite{goldsmith1}.
Furthermore, the incorporation of deformations of the standard $BF$ action, such
as those studied in \cite{me}, would produce canonical versions of the 
topological invariants obtained in \cite{crm}.

\subsection*{Acknowledgements}

The author would like to thank G. Semenoff for hospitality at the University
of British Columbia, where this work was completed. Part of this work was
carried out during the PIms/APCTP/CRM Workshop ``Particles, Fields and
Strings '99" which was held at the University of British Columbia during the
summer of 1999. The author thanks the organisors and participants for having
provided a stimulating environment in which to work. This work was supported
in part by the Natural Sciences and Engineering Research Council of Canada.


\vfill
\newpage

\begin{thebibliography}{99}

\baselineskip=12pt

\bibitem{schwarz} A.S. Schwarz, Lett. Math. Phys. {\bf 2} (1978) 247; Commun.
Math. Phys. {\bf 67} (1979) 1.

\bibitem{horo} G.T. Horowitz, Commun. Math. Phys. {\bf 125} (1989) 417.

\bibitem{bt} M. Blau and G. Thompson, Ann. Phys. {\bf 205} (1991) 130.

\bibitem{gk} J. Gegenberg and G. Kunstatter, Ann. Phys. {\bf 231} (1994) 270.

\bibitem{hs} G.T. Horowitz and M. Srednicki, Commun. Math. Phys. {\bf 130} (1990) 83.

\bibitem{hol} M.J. Bowick, S.B. Giddings, J.A. Harvey, G.T. Horowitz and A.
Strominger, Phys. Rev. Lett. {\bf 61} (1988) 2823;\\ X. Fustero, R. Gambini
and A. Trias, Phys. Rev. Lett. {\bf 62} (1989) 1964;\\ J.A. Harvey and
J.T. Liu, Phys. Lett. {\bf B240} (1990) 369;\\ B. Harms and Y. Leblanc,
Phys. Rev. {\bf D45} (1992) 2880;\\ M.I. Polikarpov, U.J. Wiese and
M.A. Zubkov, Phys. Lett. {\bf B309} (1993) 133;\\ M. Sato and S. Yahikozawa,
Nucl. Phys. {\bf B436} (1995) 100;\\ E.T. Akhmedov, M.N. Chernodub, M.I. Polikarpov
and M.A. Zubkov, Phys. Rev. {\bf D53} (1996) 2087;\\ H. Fort and
R. Gambini, Phys. Lett. {\bf B372} (1996) 226; Phys. Rev. {\bf D54} (1996)
1778;\\ E.T. Akhmedov, M.N. Chernodub and M.I. Polikarpov, JETP Lett. {\bf 67}
(1998) 389.

\bibitem{YM} K.-I. Izawa, Progr. Theor. Phys. {\bf 90} (1993) 911;\\ A.S. Cattaneo,
P. Cotta-Ramusino, A. Gamba and M. Martellini, Phys. Lett. {\bf B355}
(1995) 245;\\ H. Reinhardt, in: {\it Quark Confinement and the Hadron Spectrum
II}, eds. N. Brambilla and G.M. Prosperi (World Scientific, 1997);\\ 
F. Fucito, M. Martellini and M. Zeni, Nucl. Phys. {\bf B496} (1997) 259;\\
A.S. Cattaneo, P. Cotta-Ramusino, F. Fucito, M. Martellini, M. Rinaldi,
A. Tanzini and M. Zeni, Commun. Math. Phys. {\bf 197} (1998) 571;\\
K.-I. Kondo, Phys. Rev. {\bf D58} (1998) 105019.

\bibitem{qg} A.H. Chamseddine and D. Wyler, Nucl. Phys. {\bf B340} (1990) 595;\\
D. Birmingham, R. Gibbs and S. Mokhtari, Phys. Lett. {\bf B263} (1991) 176;\\
C.G. Callan, S.B. Giddings, J.A. Harvey and A. Strominger, Phys. Rev. 
{\bf D45} (1992) 1005;\\ H. Verlinde, in: {\it String Theory and Quantum Gravity
'91}, eds. J.A. Harvey, R. Iengo, K.S. Narain, S. Randjbar-Daemi and 
H. Verlinde (World Scientific, 1992), p. 178;\\ G. Grignani and G. Nardelli, Phys.
Rev. {\bf D45} (1992) 2719; Nucl. Phys. {\bf B412} (1994) 320;\\
D. Cangemi and R. Jackiw, Ann. Phys. {\bf 225} (1993) 229; Phys. Lett. {\bf B229}
(1993) 24;\\ M. Abe and N. Nakanishi, Progr. Theor. Phys. {\bf 89} (1993) 501;\\
I. Oda and S. Yahikozawa, Class. Quant. Grav. {\bf 11} (1994) 2653;\\
J.P. Lupi, A. Restuccia and J. Stephany, Phys. Rev. {\bf D54} (1996) 3861;\\
V. Husain and S. Major, Nucl. Phys. {\bf B500} (1997) 381;\\
J.-H. Lee and O.K. Pashaev, J. Math. Phys. {\bf 39} (1998) 102;\\
L. Friedel and K. Krasnov, Class. Quant. Grav. {\bf 16} (1999) 351.

\bibitem{baez} J.C. Baez, {\it An Introduction to Spin Foam Models of
Quantum Gravity and $BF$ Theory}, gr-qc/9905087.

\bibitem{tqft} D. Birmingham, M. Blau, M. Rakowski and G. Thompson,
Phys. Rep. {\bf 209} (1991) 129.

\bibitem{braids} E. Fadell and L. Neuwirth, Math. Scand. {\bf 10} (1962) 111;\\
R.H. Fox and L. Neuwirth, Math. Scand. {\bf 10} (1962) 119;\\
J.S. Birman, Comm. Pure. Appl. Math. {\bf 22} (1968) 41.

\bibitem{ladeg} Y. Ladegaillerie, Bull. Soc. Math. {\bf 100} (1976) 255.

\bibitem{dahm} D.M. Dahm, {\it A Generalization of Braid Theory}, Ph.D. Thesis,
Princeton University (1961), unpublished.

\bibitem{goldsmith} D.L. Goldsmith, Michigan Math. J. {\bf 28} (1981) 3.

\bibitem{crm} P. Cotta-Ramusino and M. Martellini, in: {\it Knots and Quantum
Gravity}, ed. J.C. Baez (Oxford University Press, 1994);\\
A.S. Cattaneo, P. Cotta-Ramusino and M. Martellini, Nucl. Phys. {\bf
B436} (1995) 355;\\ A.S. Cattaneo, P. Cotta-Ramusino, J. Fr\"ohlich and
M. Martellini, J. Math. Phys. {\bf 36} (1995) 6137;\\
A.S. Cattaneo, J. Math. Phys. {\bf 37} (1996) 3664; Commun. Math. Phys.
{\bf 189} (1997) 795.

\bibitem{ccr} A.S. Cattaneo, P. Cotta-Ramusino and M. Rinaldi, Commun. Math.
Phys. {\bf 204} (1999) 493.

\bibitem{djt} G.V. Dunne, R. Jackiw and C.A. Trugenberger, Ann. Phys. {\bf
194} (1989) 197;\\ I.I. Kogan, Comm. Nucl. Part. Phys. {\bf 19} (1990) 305.

\bibitem{bosnair} M. Bos and V.P. Nair, Phys. Lett. {\bf B223} (1989) 61.

\bibitem{bs} M. Bergeron and G.W. Semenoff, Ann. Phys. {\bf 245} (1996) 1.

\bibitem{bmosss} A.P. Balachandran, W.D. McGlinn, L. O'Raifeartaigh, S. Sen,
R.D. Sorkin and A.M. Srivastava, Mod. Phys. Lett. {\bf A7} (1992) 1427;\\
C. Aneziris, Mod. Phys. Lett. {\bf A7} (1992) 3789;\\
A.P. Balachandran and P. Teotonio-Sobrinho, Int. J. Mod. Phys. {\bf A9}
(1994) 1569.

\bibitem{abks} C. Aneziris, A.P. Balachandran, L.H. Kauffman and
A.M. Srivastava, Int. J. Mod. Phys. {\bf A6} (1991) 2519.

\bibitem{bss} M. Bergeron, G.W. Semenoff and R.J. Szabo, Nucl. Phys. {\bf B437}
(1995) 695.

\bibitem{me} R.J. Szabo, Nucl. Phys. {\bf B531} (1998) 525.

\bibitem{cmr} M.I. Caicedo, I. Mart\'{\i}n and A. Restuccia, {\it On the
Geometry of Antisymmetric Fields}, hep-th/9711122.

\bibitem{newbf} M.I. Caicedo and A. Restuccia, Class. Quant. Grav.
{\bf 15} (1998) 3749.

\bibitem{cechcs} O. Alvarez, Commun. Math. Phys. {\bf 100} (1985) 279;\\ A.P.
Polychronakos, Nucl. Phys. {\bf B281} (1987) 241.

\bibitem{fnn} P. Freund and R. Nepomechie, Nucl. Phys. {\bf B199} (1982) 482;\\
R. Nepomechie, Phys. Rev. {\bf D31} (1985) 1921.

\bibitem{brylinski} J.-L. Brylinski, {\it Loop Spaces, Characteristic Classes
and Geometric Quantization} (Birkh\"auser, 1992).

\bibitem{botttu} R. Bott and L.W. Tu, {\it Differential Forms in Algebraic
Topology}, (Springer-Verlag, New York, 1986).

\bibitem{cmm} A. Carey, J. Mickelsson and M. Murray, Commun. Math. Phys.
{\bf 183} (1997) 707; hep-th/9711133.

\bibitem{kalk} J. Kalkkinen, J. High Energy Phys. {\bf 9907} (1999) 002.

\bibitem{mumford} D. Mumford, {\it Tata Lectures on Theta} (Birkh\"auser, Basel, 1983).

\bibitem{goldsmith1} D.L. Goldsmith, Bull. Amer. Math. Soc. {\bf 80} (1974) 62;
Math. Scand. {\bf 50} (1982) 167.

\bibitem{tze} A.M. Polyakov, Mod. Phys. Lett. {\bf A3} (1988) 325;\\
C.-H. Tze, Int. J. Mod. Phys. {\bf A3} (1988) 1959;\\
C.-H. Tze and S. Nam, Ann. Phys. {\bf 193} (1989) 419;\\
E. Witten, Commun. Math. Phys. {\bf 121} (1989) 351.

\bibitem{rolfsen} D. Rolfsen, {\it Knots and Links} (Publish or Perish, 1976);\\
D. Eliezer and G.W. Semenoff, Phys. Lett. {\bf B266} (1991) 375.

\bibitem{balss} A.P. Balachandran, A. Stern and B.-S. Skagerstam, Phys. Rev.
{\bf D20} (1979) 439.

\end{thebibliography}
\end{document}